\documentclass[9pt,twocolumn]{extarticle}
\usepackage[letterpaper,textwidth=42pc,columnsep=2pc,top=0.75in,bottom=0.75in]{geometry}
\usepackage[T1]{fontenc}
\usepackage{mathptmx}
\usepackage{graphicx}
\usepackage{caption}
\usepackage{cite}
\usepackage{booktabs}
\usepackage{array}
\usepackage{amsmath,amssymb}
\usepackage{xcolor}
\usepackage{tikz}
\usetikzlibrary{arrows.meta,positioning,fit,calc}
\usepackage[colorlinks,linkcolor=blue,citecolor=blue,urlcolor=blue]{hyperref}
\usepackage[capitalise,noabbrev]{cleveref}

\newcommand{\xspc}{$x$SPC}
\newcommand{\sspc}{$s$SPC}

\title{Learning Where to Start: Flow Matching for Graph Layout Crossing Minimization}
\author{Yifan Hu$^{1}$, Timo Brand$^{2}$, Xuefeng Li$^{1}$, Jacob Miller$^{2}$ and Stephen Kobourov$^{2}$\\[0.5ex]
{\normalsize $^1$Northeastern University, USA \quad $^2$Technical University of Munich, Germany}}
\date{}

\begin{document}
\twocolumn[{\maketitle\vspace{-1.5em}\begin{center}\centering
  \includegraphics[width=0.7\linewidth]{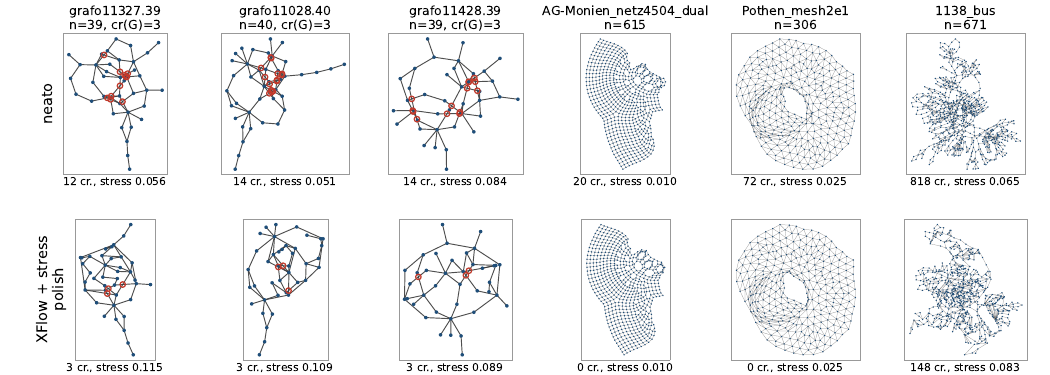}
  \captionof{figure}{Example drawings: three Rome test graphs (left; $\mathrm{cr}(G)$ is the exact crossing number) and
  three SuiteSparse graphs (right). Top row: neato. Bottom row: XFlow-cross with stress polish. Red circles mark crossings on the Rome
  graphs.}%
  \label{fig:teaser}\end{center}\vspace{1em}}]
\begin{abstract}
Edge crossings are among the layout properties that most affect how well people read node-link diagrams, but computing low-crossing layouts is difficult. Unlike other layout properties such as stress or spring-electrical energy, the number of edge crossings is not differentiable. Differentiable surrogates have been proposed and combined with gradient descent to improve layouts, but they are local methods: the layouts they find depend on where they start. The usual starting points, e.g., low-stress layouts, are not diverse. In this work, we propose the generative flow-matching model XFlow to learn a more diverse and effective set of starting layouts. First, we show that XFlow can learn to generate layouts with low stress and low spring-electrical energy, even on large and out-of-domain graphs. We then show that learning to obtain layouts with few crossings is harder, but with consistent anchored targets and expert iteration on the model's own refined samples, XFlow excels at this task: it outperforms state-of-the-art methods for crossing minimization and scales to graphs with thousands of nodes.

\end{abstract}
\section{Introduction}

Node-link diagrams are the most common way to visualize relational data, and how readable they are depends strongly on
their layout. Among the readability criteria studied, edge crossings have one of the strongest effects on how well people
read a diagram~\cite{Purchase_1997}, and people prefer drawings with fewer crossings and less
stress~\cite{Chimani_2014_stress}. Minimizing crossings is, however, hard. Computing the crossing number,
defined as the minimum number of edge crossings over all possible drawings of the graph, is
NP-hard~\cite{garey1983crossing}, and so is its straight-line variant~\cite{bienstock1991}. Exact methods can certify
the crossing number of graphs with up to about a hundred nodes~\cite{chimani2016ilp}; larger graphs need heuristics.

Since crossing is not a differentiable function of the layout, a number of differentiable surrogates have been proposed~\cite{ahmed2022multicriteria,ji2026_datafly} to be used with gradient descent. These methods can be easily
combined
with other criteria such as stress, and scale to graphs with thousands of nodes. Their weakness is that they are
local: the layout they reach depends on the layout they start from. For example they may start from a stress-minimized
 layout,
and the standard way to explore different starting points is to restart from perturbed copies of it. However, we
found that this explores little -- with the recently introduced SigmoidX
surrogate~\cite{ji2026_datafly}, gradient descent from neato~\cite{neato} removes about 45\% of the
crossings for Rome graphs (average crossings from 57.2 to 31.2 per graph), but 16 perturbed restarts only lower the average further to 27.7,
and 16 random starts do far worse (52.0; all from \cref{tab:rome}). \emph{Where to start} is a bottleneck.

\begin{figure}[htbp]
\centering
\resizebox{\columnwidth}{!}{%
\begin{tikzpicture}[font=\scriptsize, >=Stealth,
  box/.style={draw, rounded corners=2pt, align=center, minimum height=8mm, inner sep=2pt, fill=#1},
  box/.default={black!4}]
\node[box] (g) {graph $G$\\[-1pt]+ noise};
\node[box=blue!8, right=3mm of g] (m) {XFlow\\8 steps};
\node[box, right=3mm of m] (s) {$k$ layouts};
\node[box=orange!12, right=3mm of s] (d) {SigmoidX GD\\100--400 it.};
\node[box, right=3mm of d] (b) {fewest\\crossings};
\draw[->] (g) -- (m); \draw[->] (m) -- (s); \draw[->] (s) -- (d); \draw[->] (d) -- (b);
\draw[->, dashed] (b.south) -- ++(0,-4mm) -| node[pos=0.25, fill=white, inner sep=1pt]{inference: re-noise best} (m.south);
\draw[->] (b.north) -- ++(0,4mm) -| node[pos=0.25, fill=white, inner sep=1pt]{training: best become targets} (m.north);
\end{tikzpicture}}
\vspace{-0.5cm}
\caption{The XFlow-cross pipeline. The model proposes $k$ layouts, SigmoidX gradient descent refines each, and the layout
with the fewest crossings is kept. During training, the kept layouts become the model's new targets (expert iteration,
\cref{sec:xtrain}). At inference on large graphs, re-noising chains feed the best layout back to the model
(\cref{sec:renoise}).}
\label{fig:pipeline}
\vspace{-0.4cm}
\end{figure}
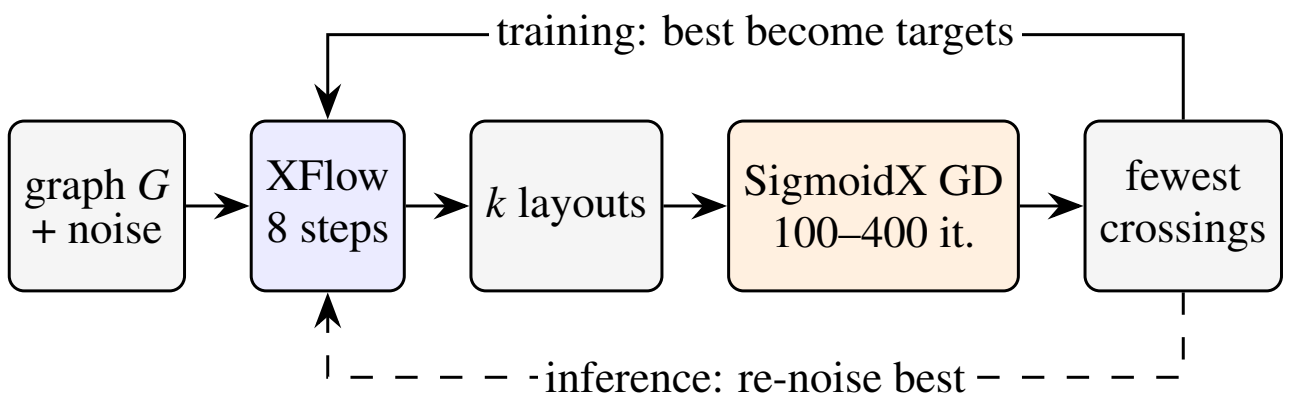

Could we use generative models to create a better kind of starting layout? Diffusion and flow-matching
models~\cite{Ho_2020_denoising_diffusion,lipman2023flow,liu2023rectified} learn a data distribution from examples,
can be trained to favor outputs with a chosen property, and produce a novel result every time they are run from random noise.
This suggests a division of labor: a learned model
proposes diverse layouts that are already close to good drawings, and gradient descent refines them locally
(\cref{fig:pipeline}). Whether this would work is not clear. The model must first learn to draw graphs at all, then learn
to draw them with few crossings, and gradient descent from its samples
must beat classical restarts at the same compute.

A DDPM~\cite{Ho_2020_denoising_diffusion} model, DiffGraph~\cite{li2026diffgraph}, was
recently applied to graph layout, but it needs many denoising steps (200) per layout.
We instead use flow matching~\cite{lipman2023flow,liu2023rectified}, whose nearly straight paths from noise to data
allow sampling in a few steps (8 or 32 in our pipeline).
We study the three hurdles above as research questions:

\textbf{RQ1.} Can a flow-matching model learn classical layouts, such as stress and spring-electrical layouts,
quickly and beyond its training distribution, and what does it learn?

\textbf{RQ2.} Can it learn to lay out graphs with few crossings?

\textbf{RQ3.} Using its samples as starting layouts, followed by gradient-descent-based crossing minimization, 
could we get layouts with fewer crossings, at matched compute, in and out of
domain, vs. state-of-the-art algorithms?

Our main contributions, which answer the three research questions, are as follows (\cref{fig:teaser} shows example drawings):
\begin{itemize}
\item \emph{RQ1 (\cref{sec:rq1}).} We introduce XFlow (X: node coordinates), to our knowledge the first
  generative approach to graph drawing based on flow matching. %
  On Rome graphs, XFlow matches neato's stress
  and sfdp's energy in 8--32 sampling steps, whereas prior diffusion layout models use 200. When trained on diverse
  graphs, XFlow remains competitive on an external benchmark of unseen graphs.
  Further analysis finds that XFlow decides the global
  layout early rather than descending the stress gradient.
\item \emph{RQ2 (\cref{sec:rq2}).} We show that naively training XFlow to imitate near-optimal crossing layouts fails. We
offer an insight into what makes a flow-matching model learn crossing minimization:
  \emph{consistent} target layouts. To further enhance performance, we
  augment the flow-matching loss with a SigmoidX term, and introduce expert
  iteration %
  on the model's own samples.
\item \emph{RQ3 (\cref{sec:rq3}).} Our crossing-reduction model produces starting layouts that make
  gradient-based crossing minimization much stronger than classical restarts at equal compute. On Rome, it nearly
  doubles the share of graphs drawn with the exact crossing number over the best baseline, S512, and beats SmartGD on
  959 of 1,000 test graphs and ties the rest; on the 91-graph benchmark~\cite{radermacher2019geometric}, it
  beats neato restarts on all graphs and improves on the best published results on all 16 named graphs. We will
  open-source our models and datasets.
\end{itemize}

\section{Related Work}

Since 1963~\cite{tutte_1963}, numerous graph drawing algorithms have been proposed. A widely used class of methods for
generating straight-line drawings relies on physical models that minimize the energy of a system. Examples include
stress models, which minimize the stress of the system%
~\cite{neato,kamada_kawai_1989,zheng-gd2}, and force-directed systems of springs and electrical charges%
~\cite{Eades_1984,Fruchterman_Reingold_1991,kobourov_2013,sfdp}. We use neato~\cite{neato} and sfdp~\cite{sfdp} 
from Graphviz~\cite{ellson2001graphviz}, two popular algorithms, as training
targets and baselines.

\subsection{Layout algorithms for optimizing specific metrics}

A number of layout methods target specific aesthetic metrics, such as edge crossings, or a combination with
stress~\cite{spx}. SGD$^2$~\cite{ahmed2022multicriteria} can jointly optimize multiple readability
criteria, as long as each can be described by a differentiable function; for edge crossings it trains a neural
network as a proxy. A recently proposed alternative, SigmoidX~\cite{ji2026_datafly}, 
is a simpler differentiable surrogate for edge crossings (\cref{sec:sigmoidx}); it achieves significantly fewer
crossings than SGD$^2$ and SmartGD~\cite{wang_2023_smartgd} when used to find optimal projections of
high-dimensional layouts to 2D. Gradient-based methods like these are fast and flexible,
but they converge to local optima that depend on the starting layout.

Of interest to the algorithmic community are algorithms that optimize crossings alone, possibly at the
expense of other criteria. %
An ILP proof system~\cite{chimani2016ilp} computes exact crossing numbers,
with curved edges allowed, for most Rome graphs; for straight-line drawings these are lower bounds. Vertex
Movement~\cite{xing-heuristic} starts from an initial layout, %
and moves each vertex to a
crossing-minimal position; it is a strong heuristic but has high computational complexity. Radermacher and
Rutter~\cite{radermacher2019geometric} make this scalable by sampling candidate positions, and publish results on a
benchmark of 91 graphs that we use as our out-of-domain test.

\subsection{Learning-based graph drawing}

Deep learning methods have been proposed for graph drawing, learning from example layouts~\cite{deep-drawing,kwon-ma-2020} or minimizing layout metrics directly~\cite{dnn,deepgd}.
Edge crossings are
not differentiable, so Tiezzi et al.~\cite{gnn-gd} train a network as a differentiable crossing surrogate
that supervises a drawing network. 
SmartGD~\cite{wang_2023_smartgd}, a generative adversarial network, learns from ``good'' layouts, where ``good'' can be defined by non-differentiable metrics such as edge crossings. Its self-challenging mechanism replaces a training example whenever the generator draws a better layout; our expert iteration (\cref{sec:xtrain}) builds on the same idea.
These methods produce one layout per graph and work best on small graphs. 
DiffGraph~\cite{li2026diffgraph} is a denoising diffusion model with an EGNN~\cite{satorras_21_equivariant_gcn} denoiser, trained on about 300,000 layouts from six algorithms on node-pruned Rome and SuiteSparse graphs, and evaluated on held-out graphs of the same collection by how closely its layouts' aesthetic metrics match the ground truth. XFlow is also equivariant, but is a few-step flow-matching model that adds all-pairs attention with shortest-path biases, and we test it on unseen kinds of graphs (\cref{sec:rq1}). Our goal also differs: rather than imitating a mixture of classic layout styles, we use generation as a search for drawings with few crossings.

\subsection{Generative models as search}

Flow matching~\cite{lipman2023flow,liu2023rectified} trains a model to transport noise to data along nearly straight paths, allowing few-step sampling; ELF~\cite{hu2026elf} extends it to language models, and other works use it to generate graph structure~\cite{eijkelboom2024vfm,lu2026digfm}. We use it instead to draw a given graph. Noising a solution partway and denoising it again yields a nearby variant~\cite{meng2022sdedit}, and repeating this from the current best helps search-like tasks~\cite{miele2026sbd}; our re-noising chains (\cref{sec:renoise}) use this to explore around good layouts.
In expert iteration~\cite{anthony2017expert}, a model guides a search and is then trained on the search's results. In our case, the model supplies starting layouts, SigmoidX refinement is the search, and the refined drawings become the new training targets. SmartGD's self-challenging mechanism is related but has no separate search: it keeps the generator's own drawing when that drawing is better.

\section{Learning Where to Start}
\label{sec:method}

We propose using a generative model to determine where gradient descent to minimize edge crossings should start (\Cref{fig:pipeline}). We use a flow-matching model trained on good drawings to transform random noise into a new layout of the input graph, providing $k$ diverse and promising starting layouts. We then improve each layout with gradient descent using the SigmoidX crossing surrogate, and select the one with the fewest crossings. During training, the refined layouts become new targets, teaching the model to generate starting points that refine well. At inference, re-noising chains feed the best layout found so far back into the model.

\subsection{Flow-matching layout model}
\label{sec:model}

Diffusion models generate data by learning to reverse a gradual noising process.
For example, DDPM~\cite{Ho_2020_denoising_diffusion} corrupts the data through a
sequence of increasingly noisy samples, and a neural network learns to remove a
small amount of noise at each step. Sampling requires hundreds of small
denoising steps, in part because the trajectories from noise to data are curved.
Flow matching~\cite{lipman2023flow,liu2023rectified} instead recasts generation as
transport along simple paths. A training sample $x$ is paired with noise
$z \sim \mathcal{N}(0, I)$ and interpolated along the straight path
\begin{equation}
  x_t = t\,x + (1-t)\,z, \qquad t \in [0, 1],
\end{equation}
and a neural network $v_\theta(x_t, t)$ is trained to predict the velocity
$v = \mathrm{d}x_t/\mathrm{d}t = x - z$.
\emph{Sampling} integrates the learned velocity from pure noise at $t = 0$ to a sample
at $t = 1$. Because each path is straight, $x = x_t + (1-t)\,v$, so a single
prediction already gives an estimate of the endpoint and a few large integration
steps suffice.

We follow the recipe of ELF~\cite{hu2026elf} 
and use a network to predict the clean sample $x$, rather
than the velocity $v$. The network outputs an estimate
$\hat x = f_\theta(x_t, t)$ of the clean data, from which the velocity follows as
\begin{equation}
  \hat v = \frac{\hat x - x_t}{1-t}.
  \label{eq:xpred}
\end{equation}
For graph layout, $x_v \in \mathbb{R}^2$ denotes the position of vertex $v$. At each
training step, every graph $g$ in the minibatch $B$ receives its own noise level
$t_g = \sigma(u_g)$ with $u_g \sim \mathcal{N}(0,1)$, where $\sigma$ is the logistic
sigmoid. This concentrates training on intermediate noise
levels. Each node's coordinates $x_v$ are paired with independent noise
$z_v \sim \mathcal{N}(0, I_2)$. 
The loss is the squared error of the implied velocity, averaged over
all nodes in the minibatch:
\begin{equation}
  \mathcal{L}_{\mathrm{FM}} = \frac{1}{\sum_{g\in B} n_g} \sum_{g \in B}
  \sum_{v \in g} \left[\frac{\|\hat x_v - x_v\|}{\max(1-t_g,\, 0.05)}\right]^2,
\end{equation}
where $n_g$ is the number of nodes in $g$, and the denominator is clamped at $0.05$
to keep the weight bounded as $t_g \to 1$.
For the crossing models, which are trained on both small and large graphs, we instead
average first over the nodes of each graph and then over the graphs,
so that large graphs do not dominate the loss. Training layouts are centered and rescaled, so that their coordinates
have about unit variance on average, like the noise. No further canonicalization is needed.

\emph{Inference.} We start from noise $x_{t_0} \sim \mathcal{N}(0, I)$ and take $N = 8$ or 32 steps,
$x_{t_{i+1}} = x_{t_i} + (t_{i+1} - t_i)\,\hat v(x_{t_i}, t_i)$ for $i = 0, \dots, N-1$, with $\hat v$ from
\cref{eq:xpred} and $t_i = (i/N)^2$, so $t_0 = 0$ and $t_N = 1$. The time steps are denser at the noisy end where the global shape is decided. Unless stated otherwise, we use 8 steps. A useful side effect of $x$-prediction is that at every step the model exposes a complete candidate drawing
$\hat x$, whose stress and crossings we can measure (\cref{sec:rq1}). 

\emph{Model architecture.} The denoiser is an E(2)-equivariant graph transformer~\cite{satorras_21_equivariant_gcn} with 8 layers, width 240 and
7.1M parameters.
Each node
starts with 9 features that describe its place in the graph: its random-walk return probabilities $(P^k)_{ii}$ for
$k = 1,\dots,8$, where $P = D^{-1}A$ ($D$ and $A$ the degree matrix and adjacency matrix, respectively), and $\log(1+\deg)$.
An MLP lifts them to a 240-dimensional hidden state. Each layer runs multi-head attention on the complete graph, over
all node pairs (XFlow-Pivot, below, uses a sparse graph), with attention biased by a learned function of the shortest-path distance $d$ (initially $\propto 1/d$) and by the
nodes' current distance. The hidden states get
the usual transformer update (attention output, then a feed-forward network, each with a residual connection). 
The
same attention weights also move the 2D positions $x_i \in \mathbb{R}^2$, in the form of EGNN's coordinate
update~\cite{satorras_21_equivariant_gcn},
\[
x_i \leftarrow x_i + \sum_j \sum_h a^h_{ij}\, c^h_{ij}\, \frac{x_i - x_j}{\|x_i - x_j\| + 1},
\]
where $a^h_{ij} \ge 0$ is head $h$'s attention weight, which selects the nodes $i$ responds to, and $c^h_{ij}$
is a signed step, 
computed like attention scores but without the softmax.
A variant, XFlow-Pivot, attends on a sparse graph instead: the graph's own edges plus links
between every node and $K = \max(16, 0.05n)$ max-min pivot nodes, with a tenth node feature flagging pivots, at a cost
of $O(n(K + \deg))$ per layer instead of $O(n^2)$.
Further details of the model architecture can be found in \cref{S-sec:arch} of the supplemental material.

\subsection{Refinement by SigmoidX gradient descent}
\label{sec:sigmoidx}

SigmoidX~\cite{ji2026_datafly} is a recently introduced effective proxy for edge crossings. 
We introduce it briefly here.
Let $t$ and $u$ be the intersection parameters of two edges $AB$ and $CD$, where the lines through them meet at
$A + t(B-A) = C + u(D-C)$. The edges cross exactly when $t, u \in [0, 1]$. SigmoidX replaces
this hard, non-differentiable test with sigmoid-based soft masks,
\[
M_t = \frac{\sigma_T(t)\,[1 - \sigma_T(t - 1)]}{M_\mathrm{peak}}, \qquad
M_u = \frac{\sigma_T(u)\,[1 - \sigma_T(u - 1)]}{M_\mathrm{peak}},
\]
where $\sigma_T(u) = 1/(1 + e^{-Tu})$ is the sigmoid function with temperature $T$ and
$M_\mathrm{peak} = \sigma_T(0.5)\,[1 - \sigma_T(-0.5)]$ normalizes the peak value to 1. The differentiable crossing
probability $P(\mathrm{crossing}) = M_t M_u$ approaches 1 when the edges intersect and decreases smoothly toward 0
otherwise. Summed over all pairs of edges that share no endpoint, it gives a smooth crossing count
$\mathrm{SigmoidX}_T(x)$ of a layout $x$. 

We refine layouts with SigmoidX: after scaling a layout to graph-distance units by its optimal stress scale, Adam (learning rate 0.3) runs 100 or 400 iterations while $T$ is annealed geometrically from 2 to 60. We keep the iterate with the fewest exact crossings, so refinement never adds crossings; overlapping and touching edges count as crossings, to avoid degenerate layouts.

\subsection{Training targets with few crossings}
\label{sec:learnx}

The XFlow model learns by imitation, so each training graph needs a target drawing with few crossings. Exact crossing numbers are known for most Rome graphs~\cite{chimani2016ilp}, but the drawings are not public, so we build three sets of targets: Targets A and B use a vertex-moving search on the Rome${}_R$ training graphs (\Cref{S-sec:vmove}), and Target C replaces the search with SigmoidX gradient descent so that it scales to larger graphs of the Mixed set.

\emph{Target A (planarization).} We use OGDF's planarization~\cite{Chimani2014OGDF}, add the missing edges back, and run a vertex-moving search. 26\% of these drawings are optimal. However, planarization can draw similar graphs very differently, which makes the targets hard to learn (\cref{sec:consistency}).

\emph{Target B (anchored, vertex search).} To make the targets consistent, we start the same search from the neato layout and from two copies with every node jittered by 5\% of the layout's spread, keep the best, and run 10 passes that lower stress without adding crossings. Fewer drawings are optimal than with Target A (19\%), but each refines the neato layout, so similar graphs get similar targets.

\emph{Target C (anchored, SigmoidX).} Vertex-moving search is slow and does not scale to large graphs, so Target C keeps the anchoring but replaces the search with gradient descent on SigmoidX: from the neato layout alone, 400 iterations (100 for graphs with more than 300 nodes). A stress polish (\cref{S-sec:polish}) then lowers stress without adding crossings; if the result's stress exceeds three times neato's, the graph keeps the neato layout as its target. This halves neato's crossings on the training graphs.

\subsection{Training for few crossings}
\label{sec:xtrain}

\emph{Crossing-aware loss.} During training, we add a crossing loss to the flow-matching loss, applied only at
low noise ($t > 0.5$):
\begin{gather*}
\mathcal{L}_{\mathrm{cross}} = \mathbb{E}\big[\mathbf{1}(t > 0.5)\,t^2\,\mathrm{SigmoidX}_{T=3}(\hat x)/\max(c_0, 1)\big],\\
\mathcal{L} = \mathcal{L}_{\mathrm{FM}} + \lambda\,\mathcal{L}_{\mathrm{cross}},
\end{gather*}
where the expectation is over the graphs in the minibatch, each with its own $t$, and dividing by $c_0$, the
number of crossings of the graph's neato layout, puts graphs with a few and with millions of crossings on one scale.
It acts on the predicted clean layout $\hat x$.
On graphs with more than 200,000 pairs of
edges, SigmoidX is evaluated on a uniform sample of 200,000 pairs and scaled up.
The weight $\lambda$ is adaptive. It starts at 0, and after each batch we add $0.1$ times the mean relative excess
$(c(\hat x) - c^\ast)/\max(c^\ast, 1)$ of the true crossings of $\hat x$ over those of the target layout, $c^\ast$. We clip $\lambda$ to $[0, 30]$.

\emph{Expert iteration.} The flow-matching models learn from targets that may not be optimal, and the models can
surpass them.
We therefore update the targets with the model's best refined drawings and fine-tune the model on them, a form of expert iteration~\cite{anthony2017expert}. In each round, the model draws 16 samples per training graph and SigmoidX refines each for 400 iterations; the refined sample with the fewest crossings (ties: lower stress) replaces the target if it has fewer crossings, and is then stress-polished (\cref{S-sec:polish}). The new targets come from the model's own choices, hence stay consistent. To reduce cost, the first three rounds update only graphs with at most 300 nodes and 1,000 edges; the fourth also updates the larger graphs, with 4 samples, 100 SigmoidX iterations and 100 polish steps. After each round, we fine-tune the model on the new targets with the same loss and learning rate $10^{-4}$, for 20K iterations in the first round and 40K later, starting from the previous round's checkpoint with the fewest crossings on held-out graphs.

\emph{Training the crossing model.} Our final crossing model, XFlow-cross (\cref{tab:models}), is trained in three
stages. First, a stress model is trained on neato layouts of  the 58K
Diverse-Aug dataset (\cref{tab:datasets}) %
(160K iterations). Second, it is fine-tuned on the Mixed set of 17K graphs 
for 20K iterations on Target C, with the crossing-aware loss; we call the resulting model XFlow-cross$^0$. Third, four rounds of
expert iteration on Mixed turn XFlow-cross$^0$ into XFlow-cross.

\begin{table}[t]
\centering\footnotesize
\caption{Trained models. FM: flow-matching loss; X: crossing-aware loss. A superscript counts rounds of expert iteration;
-D marks training on Diverse, -R crossing models trained on Rome only; XFlow-Pivot variants are trained like their all-pairs counterparts.
Sets as in \cref{tab:datasets}. $^\dagger$Trained on neato layouts of Diverse-Aug, without any validation or test graph or its variants.}
\label{tab:models}
\setlength{\tabcolsep}{3pt}
\resizebox{\columnwidth}{!}{%
\begin{tabular}{l>{\raggedright\arraybackslash}p{2.9cm}ll}
\toprule
Model & Trained on & Loss & Starts from \\
\midrule
\multicolumn{4}{l}{\emph{RQ1: learning classic layouts (\cref{sec:rq1})}} \\
XFlow-stress & neato layouts, Rome & FM & scratch \\
XFlow-stress-D & neato layouts, Diverse & FM & scratch \\
XFlow-sfdp & sfdp layouts, Rome & FM & scratch \\
XFlow-sfdp-D & sfdp layouts, Diverse & FM & scratch \\
\midrule
\multicolumn{4}{l}{\emph{RQ2: imitation on Rome (\cref{sec:rq2})}} \\
XFlow-cross-R$^0$ & Target B, Rome & FM + X & scratch \\
\midrule
\multicolumn{4}{l}{\emph{RQ2--3: the crossing model (\cref{sec:rq2-model}, \cref{sec:rq3})}} \\
XFlow-cross$^0$ & Target C, Mixed & FM + X & stress model$^\dagger$ \\
\textbf{XFlow-cross} & Target C updated by expert iteration (4 rounds), Mixed & FM + X & XFlow-cross$^0$ \\
\bottomrule
\end{tabular}%
}
\vspace{-0.4cm}
\end{table}

\subsection{Learned starts at inference}
\label{sec:renoise}

At inference the model is a source of starting layouts: each sample is refined by SigmoidX
(\cref{sec:sigmoidx}), and we keep the refined layout with the fewest crossings (ties: lower stress). Given a budget
of $k$ SigmoidX runs per graph, we spend it in one of two ways.

\emph{Independent samples (best of $k$).} Draw $k$ independent samples from noise and refine each with SigmoidX. Because every sample starts from
fresh noise, the $k$ starts are spread over the model's distribution of plausible layouts. This is in contrast to pure random restarts and perturbed restarts from the neato layout.

\emph{Re-noising chains.} Independent samples give diverse starts, but each is drawn without regard to what earlier samples found after refinement. A chain instead \emph{re-noises} the best refined layout $X$ found so far: it mixes $X$ with fresh noise to an intermediate time $t_0$,
\[
x_{t_0} = t_0\,X + (1 - t_0)\,z, \qquad z \sim \mathcal{N}(0, I),
\]
and lets the model denoise $x_{t_0}$ back to $t = 1$ before SigmoidX refines the result again. The re-noised state
keeps the coarse structure of $X$ but blurs its details, and the model then redraws those details as a plausible
layout, so the new start is a variation of the current best rather than an unrelated sample. 
Intuitively, best of $k$ explores while a chain exploits: a moderate amount of noise shifts the current best $X$,
the model repairs it into a plausible layout, and SigmoidX then lands in a different local optimum nearby.

We run $C$ chains of $R$ rounds. Round 0 of each chain is an ordinary sample; each later round re-noises the chain's
best layout so far. The answer is the best layout over all chains and rounds. 
With $C R = k$, both strategies cost the same $k$ SigmoidX runs. For $k = 16$ we use $C = R = 4$ chains and rounds
and choose $t_0 = 0.5$ based on the validation set Large-Val (\cref{S-sec:largeval}).

\section{Experimental Setup}
\label{sec:protocol}

In this section we describe the datasets, metrics, and baselines used in our experiments. XFlow models and SigmoidX
run on one NVIDIA V100 (32\,GB); neato and sfdp on one core of a Xeon E5-2680 v4. 

\subsection{Datasets}
\label{sec:datasets}
We use seven sets of graphs (\cref{tab:datasets}), detailed in \cref{S-sec:datasets} of the supplemental material. %

\begin{table}[t]
\centering\small
\caption{Graph sets. Train / val.\ / test: number of graphs in each split.}
\label{tab:datasets}
\setlength{\tabcolsep}{3pt}
\resizebox{\columnwidth}{!}{%
\begin{tabular}{llll}
\toprule
Set &  train / val.\ / test & Nodes & Used for \\
\midrule
Rome & 10,000 / 531 / 1,000 & 10--110 & in-domain training and test \\
Diverse & 5,259 / 105 / 218 & 20--1,000 & XFlow-stress-D, XFlow-sfdp-D \\
Diverse-Aug & ~58,000 / -- / -- & 10--2,000 & XFlow-cross, stage 1 (stress) \\
Mixed & 17,100 / -- / -- & 10--1,000 & XFlow-cross, stages 2--3 (crossings) \\
Large-Val & -- / 100 / -- & 301--1,000 & large-graph settings\\
Large-Test & -- / -- / 78 & 309--999 & large-graph test\\
RR & -- / -- / 91 & 102--7,920 & out-of-domain test only \\
\bottomrule
\end{tabular}%
}
\vspace{-0.4cm}
\end{table}

\emph{Rome.} The Rome dataset contains 11,534 graphs with up to 110 nodes, 
10,756 of which have known exact crossing numbers~\cite{chimani2016ilp}. 
For our crossing model and \cref{sec:rq3}, we use the split released by SmartGD~\cite{wang_2023_smartgd}
(10,000 training, 531 validation and 1,000 test graphs), which we call Rome${}_S$, so that we can compare with SmartGD on its own test graphs.
Of these test graphs, 783 have a known exact crossing number.
The analyses in \cref{sec:rq1,sec:rq2} use a random split, Rome${}_R$, of the 10,756 graphs with known crossing numbers
(9,656 / 100 / 1,000; \cref{S-sec:rome}). SmartGD's test graphs are larger than Rome's average (mean 71 nodes, 58\%
with at least 90 nodes, against 52 and 12\% over all of Rome), so neato draws 57.22 crossings per graph on them,
against 25.23 on the test graphs of Rome${}_R$. 

Despite its size, the Rome dataset 
offers limited diversity: its graphs were generated from only 112 
seed graphs~\cite{rome_graphs}. Consequently, performance on a Rome 
test split measures in-distribution generalization only.

\emph{Diverse.} To test generalization we built Diverse: 5,259 training graphs with 20--1,000 nodes (2,000 Rome
graphs, 3,000 synthetic graphs from 10 generators and 259 SuiteSparse matrices~\cite{Davis_Hu_ufl_2009}), 105
validation and 218 test graphs.

\emph{Diverse-Aug.} Following DiffGraph, we augment the Diverse training set by adding ten variants of each graph,
made by removing 5--40\% of its nodes or by growing it through node duplication. This gives about 58,000 graphs, on whose neato layouts we pretrain the stress model that XFlow-cross starts from (stage 1, \cref{sec:xtrain}).

\emph{Large-Val.} 100 Diverse graphs with more than 300 nodes, used to choose 
settings for XFlow-cross on large graphs. Diverse's validation split has only 18 such graphs, so we hold out 82 more
from its training split. The training set for XFlow-cross, Mixed (below), excludes all these.
\emph{Large-Test.} The 78 Diverse test graphs with more than 300 nodes, used only for testing.

\emph{Mixed.} The 17,100 graphs on which XFlow-cross learns crossings (stages 2--3). Its targets need SigmoidX runs on every graph, so Mixed keeps the Diverse training originals (except Large-Val) but only 4,000 of their variants, those with 300--1,000 nodes. Its Rome graphs are the Rome${}_S$ training graphs rather than Diverse's, so that no SmartGD test graph is trained on (\cref{S-sec:diverse}). No stage uses a validation or test graph or a variant of one.%

\emph{RR.} Radermacher and Rutter's benchmark~\cite{radermacher2019geometric}: 16 named real-world graphs and 75 random 3-, 6- and 9-regular graphs with 1,000 nodes. The original regular graphs are not
published, so we generated new ones (\cref{S-sec:rr}); only the 16 named graphs can be compared with
published results (\cref{S-tab:rr}). No RR graph is used for training or selection.

\subsection{Metrics}

 For a layout $X = \{x_i\}_{i \in V}$ of $G = (V, E)$, with $d_{ij}$ the shortest-path distance
between nodes $i$ and $j$, we use three metrics, following Ji et al.~\cite{ji2026_datafly}. For all three, smaller is better.

\emph{Stress} measures how well layout distances match graph distances:
\begin{equation}
\mathrm{stress} = \frac{\sum_{i<j} w_{ij}\,\big(\alpha\|x_i - x_j\| - d_{ij}\big)^2}{\sum_{i<j} w_{ij}\,d_{ij}^2},
\qquad w_{ij} = d_{ij}^{-2},
\end{equation}
where the scale $\alpha = \sum_{i<j} w_{ij} d_{ij}\|x_i - x_j\| \,/ \sum_{i<j} w_{ij}\|x_i - x_j\|^2$ minimizes the
numerator, so that the metric is scale invariant.

\emph{Spring-electrical energy} treats edges as springs and nodes as repelling charges. This is the energy that sfdp
minimizes~\cite{sfdp}:

\begin{equation}
\mathrm{energy} = \frac{1}{P}\Big(\sum_{(i,j) \in E} \frac{\|t x_i - t x_j\|^3}{3} - \sum_{i \neq j} \ln \|t x_i - t x_j\|\Big),
\end{equation}
with $P = |V|(|V|-1)$ the number of ordered node pairs, and the scale
$t = \big(P / \sum_{(i,j) \in E}\|x_i - x_j\|^3\big)^{1/3}$
that minimizes it, making the metric scale invariant. The energy can be negative.

\emph{Edge crossings} count the pairs of edges without a common endpoint that cross:
\begin{equation}
\mathrm{xing} = \sum_{\{(i,j), (k,l)\} \subseteq E,\ \{i,j\} \cap \{k,l\} = \emptyset}
\mathbb{I}\big[ x_i x_j \text{ crosses } x_k x_l\big].
\end{equation}
We count touching as crossing: an endpoint within $10^{-3}$ of the shorter edge's length from the other edge, which
covers collinear overlaps and nodes lying on edges. Without this, a layout collapsed onto a line would have no
crossings. A drawing is \emph{at optimum} when its crossings equal the graph's exact crossing number, which certifies
it as an optimal straight-line drawing.

We compare a method's value $v$ with a baseline's value $b$ by the
symmetric percentage change (SPC)~\cite{deepgd}, %
$$\mathrm{SPC} = (v - b)/\max(|v|, |b|)$$ 
\noindent averaged over graphs.
\emph{Negative values favor the method}. E.g., $-0.1$ means about 10\% better. \xspc{} and \sspc{} denote the SPC of
crossings and of stress; unless stated otherwise the baseline is neato. Paired comparisons count the graphs on which
a method has fewer (wins) or more (losses) crossings than a baseline, with a two-sided sign test.

\subsection{Baselines}
\label{sec:baselines}

We use two popular layout algorithms, neato~\cite{neato} and sfdp~\cite{sfdp} from Graphviz~\cite{ellson2001graphviz},
as baselines when checking whether XFlow can learn stress or spring-electrical energy. 

For crossing minimization, every method spends the same budget: $k$ SigmoidX runs per graph with
the same number of iterations, and the result is the run with the fewest crossings. The methods differ only in where
the runs start: 1) \emph{SigmoidX from neato}: one run from the neato layout ($k = 1$). 2) \emph{neato-restart}: one run from neato and $k - 1$ runs from perturbed copies of it, with Gaussian noise of
  0.1 mean edge lengths added to every node (0.3 on Rome, where it gives better restarts).
  3) \emph{random-restart}: $k$ runs from random Gaussian layouts with the spread of neato's layout.
4) \emph{XFlow samples}: $k$ runs from samples of an XFlow model, as best of $k$ or in re-noising chains
  (re-noise, \cref{sec:renoise}). 5) We also compare with the exact crossing numbers, with the vertex-moving search that builds our targets, and with
published results: SmartGD's on Rome and Radermacher and Rutter's on RR. We report batched GPU time per graph. 

\section{Learning Classical Layouts (RQ1)}
\label{sec:rq1}

We first understand whether a flow-matching model can learn classical layouts, and if so, how it gets there.

\subsection{Learning stress on Rome, in domain}
\label{sec:rq1rome}

\begin{table}[htbp]
\centering\small
\caption{Stress models on 1,000 unseen Rome graphs: \sspc{} against neato (negative is better), averaged over 5 samples
per graph, at 32 and 8 sampling steps. Models trained on 5,000 Rome graphs ($^\ast$1,000).}
\label{tab:stress}
\begin{tabular}{lcc}
\toprule
Model & \multicolumn{2}{c}{\sspc\,$\downarrow$} \\
 & 32 steps & 8 steps \\
\midrule
GNN, messages along graph edges$^\ast$ & +0.082 & +0.130 \\
GNN, all pairs, messages weighted $1/d$ & +0.004 & +0.031 \\
Graph transformer, graph edges only & +0.177 & +0.225 \\
Graph transformer, all pairs, PCA-aligned & $-$0.007 & +0.030 \\
\textbf{XFlow: equivariant, all pairs} & \textbf{$-$0.015} & \textbf{+0.020} \\
XFlow-Pivot: equivariant, edges + pivots & $-$0.012 & +0.045 \\
\bottomrule
\end{tabular}
\vspace{-0.2cm}
\end{table}

We investigated different configurations of graph convolutional networks (GCNs), including whether using only graph edges for message passing is sufficient. As shown in \Cref{tab:stress}, the model matches neato's stress on unseen Rome graphs when each node can attend to distant nodes, either through all-pairs attention or via pivots. In contrast, models that pass messages only along graph edges fall short. XFlow slightly outperforms neato at 32 steps (\sspc{} $-0.015$) and comes close at 8 steps ($+0.020$). Rotation equivariance also outperforms PCA-based target alignment at both step counts. We therefore adopt rotation-equivariant graph transformers operating on either the complete graph or the pivot graph as our default configuration for the remainder of the experiments.

\subsection{Out of domain: learning stress and spring-electrical energy on Diverse} 

\label{sec:rq1ood}

While XFlow-stress matches neato on Rome (\cref{tab:stress}), we have to be extra careful: 
training and testing on Rome graphs is not a fair test of generalization. 
Not only is it in-domain on a set of small graphs ($\le$110 nodes),
but we also have to be aware of the fact that the Rome graphs were derived from 112 seed
graphs~\cite{rome_graphs}.

Indeed, when tested out of domain, its \sspc{} is +0.397 on the Diverse test split
and +0.450 on RR (\cref{tab:ood}, first row). So the model learned the 
Rome distribution rather than the neato algorithm. To test the generalizability of XFlow
rigorously, we need to train on a more varied set of graphs, Diverse (\cref{sec:datasets}), 
including larger ones, with neato or sfdp layouts as targets, 
and test on Rome, on the Diverse test
split, and out of domain on RR. The resulting models are 
XFlow-stress-D, XFlow-Pivot-stress-D, XFlow-sfdp-D and XFlow-Pivot-sfdp-D.

\begin{table}[htbp]
\vspace{-0.3cm}
\centering\small
\caption{Models trained on Diverse (160K iterations, 32 sampling steps), and, in the first row, XFlow-stress
trained on Rome only. SPC against the algorithm each model imitates: \sspc{} against neato for stress models, energy SPC against sfdp for sfdp models (negative is better). Rome:
the 1,000 test graphs of \cref{tab:stress}. RR: 91 graphs of kinds not seen in training.
}
\label{tab:ood}
\setlength{\tabcolsep}{4pt}
\begin{tabular}{llccc}
\toprule
Model & Against & Rome\,$\downarrow$ & Diverse test\,$\downarrow$ & RR\,$\downarrow$ \\
\midrule
XFlow-stress (Rome only) & neato & $-$0.015 & +0.397 & +0.450 \\
\midrule
XFlow-stress-D & neato & \textbf{+0.036} & \textbf{+0.073} & \textbf{+0.024} \\
XFlow-Pivot-stress-D & neato & +0.057 & +0.089 & +0.031 \\
XFlow-sfdp-D & sfdp & +0.015 & +0.060 & +0.071 \\
XFlow-Pivot-sfdp-D & sfdp & \textbf{+0.009} & \textbf{+0.036} & \textbf{+0.020} \\
\bottomrule
\end{tabular}
\vspace{-0.2cm}
\end{table}

\emph{Stress.}  As seen in \cref{tab:ood}, the models trained on Diverse
come within a few percent of their target
algorithms on RR: +0.024 for XFlow-stress-D, against +0.450 for the Rome-trained model, 
and +0.020 for
XFlow-Pivot-sfdp-D. The Diverse test split, 
although drawn from the training distribution, is harder than Rome for
every model (+0.036 to +0.089). The price on Rome is small: +0.036 instead of $-$0.015, 
since Rome is now a minority
of the training set.

\emph{Spring-electrical energy.} The same model and recipe learn sfdp as well as neato. On Rome, XFlow-sfdp-D and
XFlow-Pivot-sfdp-D come within +0.015 and +0.009 of sfdp's energy, closer than neato's own layouts (+0.028), so the
models reproduce sfdp's drawings rather than just some good drawing. Likewise, the stress models are much closer to
neato than sfdp is (+0.036 vs.\ +0.211). For stress, all-pairs attention is slightly better on all three sets. 
For sfdp the pivot
model is better, and generalizes clearly better to RR (+0.020 vs.\ +0.071), while costing $O(n(K + \deg))$ instead
of $O(n^2)$ per layer.

\emph{Speed.} On the 16 named RR graphs XFlow-stress-D takes 150\,s in total on one GPU at 32 steps, while
neato, which has no GPU implementation, takes 360\,s on one CPU core; on a 28-thread CPU, XFlow-stress-D takes 261\,s
at 8 steps and 977\,s at 32. DiffGraph reports about
800\,ms per Rome graph with 200 denoising steps on a Titan RTX~\cite{li2026diffgraph}; run one graph at a time, our
model takes 326\,ms per Rome graph at 32 steps, the setting of \cref{tab:ood}, and 91\,ms at 8 steps (\cref{tab:stress}
shows that the quality of the two is close); batched, generation costs about 1.6\,ms per graph per step. 
Our V100 and DiffGraph's Titan RTX have similar throughput, so most of the gap comes from the
number of steps, 200 against 32 or 8: flow matching is known to need far fewer sampling steps than
DDPM~\cite{Ho_2020_denoising_diffusion}, the denoising diffusion DiffGraph uses, because its paths from noise to data are nearly
straight~\cite{lipman2023flow,liu2023rectified}.

\subsection{What the XFlow-stress model learns}
\label{sec:rq1traj}

\begin{figure*}[t]
\centering
\includegraphics[width=\linewidth]{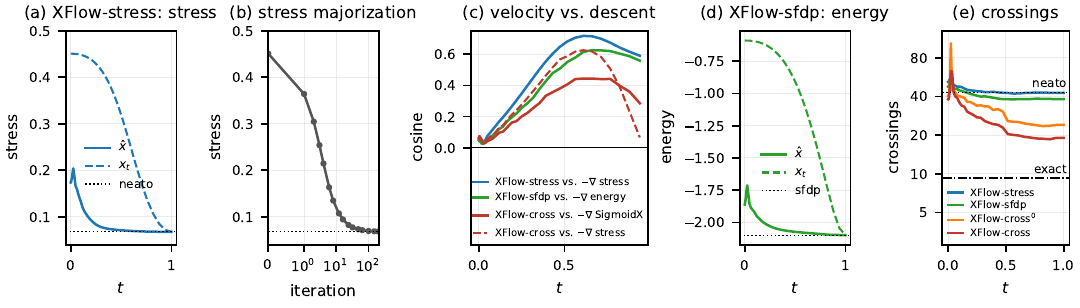}
\vspace{-0.7cm}
\caption{Along the sampling trajectory (32 steps; mean over the 84 Rome graphs that are test graphs in both Rome${}_R$ and Rome${}_S$)
XFlow-stress and XFlow-sfdp are trained on Rome, XFlow-cross$^0$ and XFlow-cross on Mixed.
(a) Stress of XFlow-stress's prediction $\hat x(x_t, t)$ (solid) and of the current state $x_t$ (dashed).
(b) Stress majorization with neato's weights from the same initial noise, per iteration. (c) Cosine between each
model's sampler velocity $(\hat x - x_t)/(1-t)$ and the negative gradient at $x_t$ of the objective it targets
(stress, spring-electrical energy, SigmoidX), and, dashed, of stress for XFlow-cross.
(d) Spring-electrical energy of XFlow-sfdp's prediction and state, with sfdp's energy. (e) Crossings of the
prediction for XFlow-cross$^0$ and XFlow-cross; XFlow-stress, XFlow-sfdp, neato's crossings and the mean exact
crossing number of these graphs are shown for reference.
The schedule is $t_i = (i/32)^2$.}
\label{fig:traj}
\vspace{-0.7cm}

\end{figure*}

XFlow generates a layout by moving random positions toward a layout in a few steps (\emph{sampling}). Does each
step simply move downhill on the objective it imitates?
 We sampled the 84 Rome graphs that are test graphs in both
Rome${}_R$ and Rome${}_S$ (\cref{sec:datasets}) with 32 steps, $t_i=(i/32)^2$, using four models: XFlow-stress and XFlow-sfdp,
trained on stress and sfdp layouts of Rome, and the crossing models XFlow-cross$^0$ and XFlow-cross (\cref{tab:models}; studied
in \cref{sec:rq2}), and recorded, at every step,
the current state $x_t$, the model's
prediction $\hat x$ of the final layout, and the sampler's velocity (\cref{fig:traj}). Two observations stand out.
First, \emph{the model predicts a good layout early}. For example, for XFlow-stress, halfway through the 32 steps
($t = 0.25$), the prediction $\hat x$ already has stress 0.084, against 0.17 at $t = 0$ and a final 0.068, equal to
neato's 0.069, while the current state $x_t$ is still mostly noise (\cref{fig:traj}a).
The remaining steps mostly
refine it. Step for step the
prediction improves at about the rate of stress majorization from the same noise (\cref{fig:traj}b), but it reaches
neato's stress in 32 steps where majorization needs about 100 iterations. Second, \emph{the sampler is not gradient
descent}. 
The cosine between its velocity and the negative stress gradient rises from about 0.06 to at most 0.72, at
$t \approx 0.6$, and then falls (\cref{fig:traj}c). Hence the model is not following the steepest-descent direction, 
which would result in a cosine of 1. XFlow-sfdp
behaves the same way with respect to its energy. 

\emph{Discussion on RQ1.} A flow-matching model learns stress and spring-electrical
layouts to the quality of the classical algorithms in a few sampling steps. Trained on Diverse, it stays within
a few percent of both algorithms on unseen kinds of graphs. During the flow-matching process, the model 
homes in on a global shape early and refines it. 
The model's velocity is somewhat, but not fully, aligned with the negative gradient of the objective it imitates.

\section{Learning to Draw with Few Crossings (RQ2)}
\label{sec:rq2}

To use a flow-matching model for drawings with few crossings, it must learn from examples of such drawings. We
first trained XFlow on the drawings with the fewest crossings we could build, Target A of \cref{sec:learnx}, and
found that it fails. Understanding why led us to Target B and then to our final crossing model, which is trained on
Target C (\cref{sec:rq2-model}).

\begin{table}[htbp]
\vspace{-0.4cm}
\centering\small
\caption{Learning to minimize crossings on the 1,000 Rome${}_R$ test graphs (32 sampling steps, 5 samples per graph, no
refinement unless stated). \xspc{} and \sspc{} are relative to neato; the exact crossing number averages 5.29~\cite{chimani2016ilp}.}
\label{tab:imitation}
\setlength{\tabcolsep}{4pt}
\resizebox{\columnwidth}{!}{%
\begin{tabular}{lccc}
\toprule
Layout & Cross. & \xspc\,$\downarrow$ & \sspc\,$\downarrow$ \\
\midrule
neato (baseline) & 25.23 & 0 & 0 \\
\midrule
\multicolumn{4}{l}{\emph{Training targets (ground truth)}} \\
\quad Target A: planarization & 11.10 & $-$0.501 & +0.469 \\
\quad Target B: anchored, vertex search & 13.54 & $-$0.442 & +0.199 \\
\midrule
\multicolumn{4}{l}{\emph{XFlow models}} \\
\quad XFlow-stress (no crossing training) & 24.82 & $-$0.024 & $-$0.012 \\
\quad XFlow-stress fine-tuned on Target A & 26.35 & +0.024 & +0.485 \\
\quad XFlow-stress fine-tuned on Target B & 20.62 & $-$0.181 & +0.166 \\
\quad XFlow-cross-R$^0$: Target B + crossing-aware loss & \textbf{10.70} & \textbf{$-$0.564} & +0.393 \\
\midrule
\multicolumn{4}{l}{\emph{Gradient descent}} \\
\quad SigmoidX from neato, 400 iterations & 13.77 & $-$0.410 & +0.483 \\
\bottomrule
\end{tabular}%
}
\vspace{-0.5cm}

\end{table}

\subsection{Naively imitating optimal crossing drawings fails}
\Cref{tab:imitation} shows the central negative result. On the test graphs, Target A has about half of neato's
crossings (11.10 vs.\ 25.23, \xspc{} $-$0.50), yet XFlow-stress fine-tuned on it
draws more crossings than neato (26.35, \xspc{} +0.024).
We conjecture that this is because 
planarization via OGDF does not have the \emph{consistency property}, that similar graphs have similar
layouts. It optimizes crossings alone, so a small change to a graph can change its whole drawing. The
flow-matching loss is a squared error, so for graphs that look alike the model learns the mean of their targets;
the mean of very different drawings is not itself a good drawing and may have many crossings.

The results on Target B support this conjecture. Target B has more crossings than Target A but is consistent: each
of its layouts is a small edit of the stress layout, and stress layouts change little when the graph changes
little, which is why XFlow-stress can learn them almost exactly (\cref{tab:stress}). XFlow-stress fine-tuned on
Target~B has 18\% fewer crossings than neato. Consistent targets are not enough, however: the model also needs
to be told to avoid crossings. Adding the crossing term of \cref{sec:xtrain} to the loss (XFlow-cross-R$^0$)
halves the crossings again, to 10.70 against neato's 25.23 (\xspc{} $-$0.56). We test the conjecture directly
next.

\begin{table}[htbp]
\centering\small
\caption{Consistency of the targets: median Procrustes distance between the drawing of a Rome graph and that of the same graph
with one added edge between nodes at graph distance 2 or 3, or drawn with another random seed (200 Rome${}_S$
test graphs; 0: identical, 1: unrelated). neato and Target C, which starts from neato, give the same drawing for
almost every seed.}
\label{tab:consistency}
\begin{tabular}{lccc}
\toprule
& \multicolumn{2}{c}{Added edge} & \\
\cmidrule(lr){2-3}
Method & distance 2 & distance 3 & Other seed \\
\midrule
neato & 0.07 & 0.13 & 0.00 \\
sfdp & 0.14 & 0.22 & 0.18 \\
Target A & 0.77 & 0.77 & 0.30 \\
Target B & 0.25 & 0.28 & 0.22 \\
Target C & 0.31 & 0.35 & 0.00 \\
\bottomrule
\end{tabular}
\vspace{-0.3cm}
\end{table}

\subsection{Validating the consistency conjecture\label{sec:consistency}}
We take 200 Rome${}_S$ test graphs and make five variants of each by adding one edge,
between two nodes at graph distance 2 (three variants) or 3 (two variants). We draw each graph and its variants
with each target method and measure how far a variant's drawing is from the drawing of the original graph, using the Procrustes distance
(the $L_2$ distance after the best rotation, reflection, translation and scaling, relative to the size of the
drawing; 0: identical, 1: unrelated). We also redraw each original graph with two other random seeds. Adding one edge changes a Target A
drawing almost completely (median 0.77, \cref{tab:consistency}), far more than redrawing the same graph with another
seed does (0.30). Targets B and C change less than half as much (0.25--0.35). 
This supports the conjecture: \emph{Target A
has fewer crossings than Target~B, but its drawings change the most between similar graphs, so XFlow cannot learn
from it.} Further details are in \cref{S-sec:consistency} of the supplemental material.

\subsection{Scaling up: training XFlow-cross}\label{sec:rq2-model}
The results so far use only the small Rome graphs, and a crossing model trained on Rome alone does not transfer: it
draws more crossings than neato on the majority of the larger RR graphs. To train a crossing model for a wider range of graphs,
XFlow-cross, we replace Target~B, whose vertex search does not scale, by Target~C, which refines the neato layout
by SigmoidX gradient descent and is about as consistent (\cref{tab:consistency}). We build Target~C for Mixed, the
17,100 Rome, Diverse and larger graphs of \cref{sec:datasets}, and train XFlow-cross on it with the crossing-aware
loss (\cref{sec:xtrain}). 

\begin{table}[htb]
\vspace{-0.3cm}
\centering\small
\caption{Training XFlow-cross (round 0: XFlow-cross$^0$; round 4: XFlow-cross). Targets: mean crossings of the training targets of the 10,000 Rome${}_S$ training graphs.
Rome${}_S$: 1,000 test graphs, 16 samples with 100 SigmoidX iterations each. 
Opt.: \% of the 783 graphs with a known exact crossing number reaching the optimum (mean 7.84~\cite{chimani2016ilp}).
Large-Test and RR: re-noising chains
(\cref{sec:renoise}). neato, restart: \xspc{} against neato and against neato-restart ($k = 16$, 100 SigmoidX iterations; \cref{sec:baselines}).}
\label{tab:iter}
\setlength{\tabcolsep}{3.5pt}
\resizebox{\columnwidth}{!}{%
\begin{tabular}{lccccccccc}
\toprule
 & Targets & \multicolumn{4}{c}{Rome${}_S$ test (including \xspc)} & \multicolumn{2}{c}{Large-Test \xspc} & \multicolumn{2}{c}{RR \xspc} \\
\cmidrule(lr){3-6}\cmidrule(lr){7-8}\cmidrule(lr){9-10}
Round & Rome${}_S$ & Cross. & Opt. & neato\,$\downarrow$ & restart\,$\downarrow$ & neato\,$\downarrow$ & restart\,$\downarrow$ & neato\,$\downarrow$ & restart\,$\downarrow$ \\
\midrule
0  & 14.9 & 22.54 & 29\% & $-$0.650 & $-$0.268 & \textbf{$-$0.521} & $-$0.203 & $-$0.427 & $-$0.045 \\
1 & 9.6 & 21.58 & 34\% & $-$0.667 & $-$0.306 & $-$0.510 & $-$0.201 & $-$0.441 & $-$0.065 \\
2 & 8.9 & 20.62 & 37\% & $-$0.684 & $-$0.340 & $-$0.518 & $-$0.215 & $-$0.451 & $-$0.081 \\
3 & 8.4 & 20.01 & 38\% & $-$0.694 & $-$0.359 & $-$0.518 & $-$0.213 & $-$0.455 & $-$0.086 \\
4  & \textbf{8.1} & \textbf{19.69} & \textbf{39\%} & \textbf{$-$0.699} & \textbf{$-$0.369} & $-$0.519 & \textbf{$-$0.221} & \textbf{$-$0.466} & \textbf{$-$0.104} \\
\bottomrule
\end{tabular}%
}
\end{table}

\emph{Expert iteration.} The model's own refined samples make better targets than those built by hand: they have
few crossings, and they are consistent because they come from the model's own choices. Expert iteration
(\cref{sec:xtrain}) uses them as new targets. As \cref{tab:iter} shows, the first round removes a third of the
crossings of the Rome targets (14.9 to 9.6) and later rounds remove fewer (8.1 after four), but every round improves
the model on Rome and RR. On Large-Test, round 4 is slightly worse than round 0 against neato ($-$0.519 vs.\ $-$0.521) but better against neato-restart ($-$0.221 vs.\ $-$0.203). The two disagree because \xspc{} averages per-graph ratios: against neato, both rounds are close to $-1$ on sparse graphs, which hides round 4's gains there, and a single graph with 3 neato crossings (2 for round 0, 7 for round 4) outweighs them. Graph by graph, round 4 has fewer crossings than round 0 on 48 graphs and more on 21.

\emph{When and how crossings are removed.} 
 The trajectories in \cref{fig:traj}e show how the crossing models differ
from XFlow-stress. The models trained on stress or spring-electrical layouts, which see no crossing information,
settle their crossings by $t \approx 0.35$--$0.45$, at about neato's level. XFlow-cross$^0$ removes some crossings
early (about 40 to 31 by $t = 0.47$) and then drops to 26 in a single step at $t \approx 0.5$, exactly where its loss
starts to apply the crossing term. After expert iteration (XFlow-cross), most of the removal happens earlier and
gradually (38 to 23 by $t = 0.47$). Even so, its velocity is more aligned with stress descent (cosine at most 0.63)
than with SigmoidX descent (at most 0.44): the model reaches low-crossing layouts by its own route, not by following
the SigmoidX surrogate.

\emph{Discussion on RQ2.} A flow-matching model can learn to draw with few crossings, but not by imitating optimal
drawings. What matters is consistency of the targets: targets anchored at the stress layout, a crossing term in the
loss, %
and expert iteration, which keeps improving both the targets
and the model. Together they bring XFlow-cross to 19.69 crossings on the Rome${}_S$ test graphs against neato's 57.22
(\xspc{} $-$0.70), with 39\% of the drawings at the exact crossing number (\cref{tab:iter}; 16 samples with 100
SigmoidX iterations each).

\section{Learned Starts for Crossing Minimization (RQ3)}
\label{sec:rq3}

We have shown that XFlow-cross draws layouts with substantially fewer crossings (\cref{sec:rq2-model}), and that
refining them with gradient descent improves them further. Gradient descent with SigmoidX is local, so what it achieves
depends on where it starts. We therefore ask: \emph{does XFlow-cross give better starting points for gradient descent
than classical layouts?} As shown in \cref{tab:rome}, on the 1,000 Rome${}_S$ test graphs, where neato
draws 57.22 crossings per graph, SigmoidX from neato reaches 31.23 ($-0.44$ \xspc), but the best of 16 runs from
random layouts only 52.01 ($-0.08$ \xspc). A good start matters: XFlow-cross without any refinement (28.28, $-0.52$ \xspc)
already draws fewer crossings than SigmoidX from neato.
We therefore use XFlow-cross as a source of starting layouts
for SigmoidX and compare these learned starts with classical ones (neato, perturbed restarts, random starts) at the
same number of SigmoidX runs and iterations, first on Rome (\cref{sec:rome}) and then on unseen and large graphs
(\cref{sec:largegraphs}).

\begin{table}[htbp]
\centering\small
\caption{Results on the 1,000 Rome${}_S$ test graphs. Each row keeps the best of $k$ layouts; It.: SigmoidX
iterations per layout (0: no refinement). GPU time per graph in ms  ($^\ast$: CPU). %
SmartGD rows: per-graph crossing counts for the same graphs from the authors~\cite{wang_2023_smartgd}. SGD${}^2$~\cite{ahmed2022multicriteria} and S512~\cite{radermacher2019geometric}: our runs on CPU (AMD EPYC 9555 Turin).
Opt.: \% of the 783 graphs with a known exact crossing number reaching the optimum (mean 7.84~\cite{chimani2016ilp}). %
neato alone has 57.22 crossings.}
\label{tab:rome}
\setlength{\tabcolsep}{3pt}
\resizebox{\columnwidth}{!}{%
\begin{tabular}{lrrrrrrr}
\toprule
Starts & $k$ & It. & ms & Cross. & Opt. & \xspc\,$\downarrow$ & \sspc\,$\downarrow$ \\
\midrule
{S512} & 1 & -- & {107$^\ast$} & 28.97 & 20\% & $-$0.53 & -- \\
SGD${}^2$[stress+xing] & 1 & -- & {43,495$^\ast$} & 35.17 & 12\% & $-$0.29 & -- \\
SmartGD[stress+xing] & 1 & -- & -- & 47.85 & 4\% & $-$0.12 & -- \\
SmartGD[xing] & 1 & -- & -- & 33.72 & 5\% & $-$0.33 & -- \\
neato & 1 & 400 & 26.0 & 31.23 & 11\% & $-$0.44 & +0.49 \\
XFlow-cross$^0$ & 1 & 400 & 38.2 & 27.52 & 16\% & $-$0.52 & +0.57 \\
XFlow-cross, no SigmoidX & 1 & 0 & 12.8 & 28.28 & 20\% & $-$0.52 & +0.56 \\
\midrule
neato-restart & 4 & 400 & 108 & 29.29 & 14\% & $-$0.50 & +0.50 \\
random-restart & 4 & 400 & 104 & 60.53 & 2\% & +0.10 & +0.81 \\
XFlow-cross$^0$ & 4 & 100 & 80 & 24.73 & 24\% & $-$0.60 & +0.54 \\
XFlow-cross & 4 & 100 & 80 & 21.66 & 32\% & $-$0.66 & +0.60 \\
\midrule
neato-restart & 16 & 400 & 432 & 27.70 & 18\% & $-$0.54 & +0.51 \\
random-restart & 16 & 400 & 432 & 52.01 & 4\% & $-$0.08 & +0.81 \\
XFlow-cross$^0$ & 16 & 100 & 318 & 22.54 & 29\% & $-$0.65 & +0.55 \\
XFlow-cross$^0$ & 16 & 400 & 611 & 21.91 & 30\% & $-$0.66 & +0.57 \\
XFlow-cross & 16 & 100 & 318 & 19.69 & 39\% & $-$0.70 & +0.60 \\
\textbf{XFlow-cross} & 16 & 400 & 611 & \textbf{19.16} & \textbf{40\%} & \textbf{$-$0.70} & +0.62 \\
\bottomrule
\end{tabular}%
}
\vspace{-0.3cm}

\end{table}

\subsection{In domain performance} 
\label{sec:rome}

\Cref{tab:rome} compares the algorithms on the Rome${}_S$ test graphs at the same number of SigmoidX runs $k$, with the
GPU time of each. Plotted against GPU time (\cref{S-sec:pareto}), XFlow-cross reaches the exact crossing number more
often than neato or random restarts at every compute budget.

\emph{Classical restarts explore little.} SigmoidX from neato (\cref{sec:baselines}) reaches the exact crossing number on 11\% of the
graphs, and sixteen perturbed restarts raise this only to 18\%. Random starts are worse than neato alone even with 16
starts (52.01 crossings against 31.23): 400 iterations do not untangle a random layout.

\emph{Learned starts are much better.} With 16 samples and 100 SigmoidX iterations, XFlow-cross has 19.69 crossings
against 27.70 for 16 neato restarts with 400 iterations, at 74\% of the GPU time, and reaches the exact crossing
number on more than twice as many graphs (39\% vs.\ 18\%). Paired over the 1,000 graphs, it is better than the
restarts on 829, worse on 1 and tied on the remaining 170. Expert iteration contributes: before it, XFlow-cross$^0$
with 100 SigmoidX iterations has 22.54 crossings, against 19.69 after it. Even a single sample of XFlow-cross with no
refinement has fewer crossings than SigmoidX from neato (28.28 vs.\ 31.23) at about half the GPU time. The price of
drawings with far fewer crossings is somewhat higher stress than the restarts.

\emph{Comparison with baselines.} SmartGD~\cite{wang_2023_smartgd},
a learned generative model, has 33.72 crossings and
reaches the exact crossing number on 5\% of the graphs (\cref{tab:rome}). SGD${}^2$~\cite{ahmed2022multicriteria}, which optimizes crossings by stochastic gradient descent, draws 35.17
crossings and reaches the exact crossing number on 12\%. Both are worse than XFlow-cross without refinement, which has 28.28 crossings and reaches the exact crossing number on 20\% of the test graphs. S512~\cite{radermacher2019geometric} is the strongest baseline with 28.97 crossings and reaches the exact crossing
number on 20\% of the graphs, on par with XFlow-cross without
refinement. The learned starts pay
off: with 16 samples and 100 SigmoidX iterations,
XFlow-cross is better than SmartGD on 959 of the test graphs, ties on 41, and is never worse, and is better than S512 on 813 graphs and worse
on 7.

\subsection{Out of domain and large graphs}
\label{sec:largegraphs}

We apply the same model, XFlow-cross, to Large-Test and  RR, which contain no graph of its
training data. On graphs this large, SigmoidX costs up to 13\,s per 100 iterations, so every start gets 100
iterations.

\begin{table}[htbp]
\centering\small
\caption{Large graphs: 16 SigmoidX runs of 100 iterations per graph. \xspc{} and
\sspc{} against plain neato; \xspc{} and wins/losses (W/L) against neato-restart (noise 0.1). Starts as in \cref{tab:rome};
re-noise: re-noising chains with $t_0 = 0.5$ (\cref{sec:renoise}).}
\label{tab:large}
\setlength{\tabcolsep}{3.5pt}
\resizebox{\columnwidth}{!}{%
\begin{tabular}{llcccc}
\toprule
 & & \multicolumn{2}{c}{vs.\ neato} & \multicolumn{2}{c}{vs.\ neato-restart} \\
\cmidrule(lr){3-4}\cmidrule(lr){5-6}
Set & Starts & \xspc\,$\downarrow$ & \sspc\,$\downarrow$ & \xspc\,$\downarrow$ & W/L \\
\midrule
Large-Test & neato-restart & $-$0.447 & +0.36 & 0.000 & -- \\
 & random-restart & +0.248 & +0.74 & +0.549 & 0/78 \\
 & XFlow-cross, best of 16 & $-$0.510 & +0.44 & $-$0.203 & 66/8 \\
 & XFlow-cross, re-noise & $-$0.519 & +0.46 & $-$0.221 & 66/7 \\
\midrule
RR (91) & neato-restart & $-$0.405 & +0.35 & 0.000 & -- \\
 & random-restart & $-$0.053 & +0.56 & +0.326 & 0/91 \\
 & XFlow-cross, best of 16 & $-$0.440 & +0.43 & $-$0.060 & 91/0 \\
 & XFlow-cross, re-noise & $-$0.466 & +0.47 & $-$0.104 & 90/1 \\
\bottomrule
\end{tabular}%
}
\vspace{-0.3cm}
\end{table}

\emph{XFlow samples beat neato restarts on large and out-of-domain graphs.} On Large-Test (\cref{tab:large}), best of 16 XFlow-cross samples
beats neato with 16 restarts on 66 of 78 graphs and loses on 8, with 20\% fewer crossings per graph on average.
On RR, it wins on all 91 graphs, with 6\% fewer on average. Re-noising chains do better on both sets:
22.1\% and 10.4\% fewer crossings than the 16 neato restarts on Large-Test and RR, respectively.
On Large-Val, on which $t_0$ was chosen, the margins are slightly larger (\cref{S-sec:largeval}).

\emph{Comparison with published results.} \Cref{S-tab:rr} compares our methods with
published results in~\cite{radermacher2019geometric} on the 16 named graphs. Our neato layouts have about as many crossings as their stress layouts (geometric mean
ratio 1.05), so the starting points are comparable. A single SigmoidX run from neato has fewer crossings than their
best configuration (S512) on 12 of the 16 graphs, 15.2\% fewer per \xspc.
Best of 16 XFlow-cross samples beats S512 on 15
graphs ($-$0.235 \xspc), and re-noising chains beat it on all 16 ($-$0.265 \xspc).

\emph{Sampling seed and steps.} Results vary little with the sampling seed, and 32 sampling steps instead of 8 improve them only slightly (\cref{S-sec:seeds}).

\emph{Discussion on RQ3.} Learned starts make gradient-based crossing minimization substantially stronger. In domain, they more than double the share of optimal drawings at equal compute and are almost never worse than neato restarts. On large and out-of-domain graphs, the same model beats restarts on most graphs, and with re-noising chains, XFlow-cross improves on the best published heuristic results on all 16 named graphs.

\section{Discussion and Limitations}

SigmoidX is a good local but weak global optimizer: from neato it removes about half of the crossings, and neither
perturbed nor random restarts improve much on that. XFlow-cross supplies starts that are both diverse and close to
good drawings, and when used with SigmoidX, excels in and out of domain, on small and large graphs.

Model-based layout algorithms are often evaluated on the same class of graphs they were trained on (e.g., Rome).
We argue that a more rigorous evaluation is needed, not least because the Rome graphs were generated from only 112
seed graphs~\cite{rome_graphs}. XFlow-stress, trained on Rome, matched neato's stress on Rome, but was 45\%
worse on the larger, structurally different RR graphs (\cref{tab:ood}). This shows the importance of
training on larger and more diverse graphs, and of \emph{testing on new graph types and sizes}. %

While crossing is an important metric, crossing minimization alone tends to increase
stress. The polish step~(\cref{S-sec:polish}) reduces stress while keeping crossings low, but the final layouts of XFlow-cross 
still have higher stress than neato. Finding the right balance between stress and crossings is an open problem, 
and we leave it to future work. 

\section{Conclusions}

This paper asks whether a generative model can be effective for crossing minimization. We first show that a
flow-matching model learns classical layouts in a few sampling steps, by its own route rather than by
following the stress gradient. Learning directly from crossing-optimal layouts, however, is not effective: we show that
to draw with few crossings, the targets must also be consistent. We therefore train the model on consistent
targets, add a crossing loss to the flow-matching loss, and further improve the targets by expert iteration on the
model's own refined drawings. Used as a source of starting layouts, the model more than doubles the share of Rome
graphs drawn with the minimum number of crossings at equal compute, outperforms
perturbed neato restarts on most large graphs, and improves on the best published heuristic results on every
named graph of an external benchmark.

\bibliographystyle{alpha}
\bibliography{refs}

\newcommand{\etalchar}[1]{$^{#1}$}
\begin{thebibliography}{DADL{\etalchar{+}}19}

\bibitem[ADLD{\etalchar{+}}22]{ahmed2022multicriteria}
Reyan Ahmed, Felice De~Luca, Sabin Devkota, Stephen Kobourov, and Mingwei Li.
\newblock Multicriteria scalable graph drawing via stochastic gradient descent,
  {$(SGD)^{2}$}.
\newblock {\em IEEE Transactions on Visualization and Computer Graphics},
  28(6):2388--2399, 2022.

\bibitem[ATB17]{anthony2017expert}
Thomas Anthony, Zheng Tian, and David Barber.
\newblock Thinking fast and slow with deep learning and tree search.
\newblock In {\em Advances in Neural Information Processing Systems},
  volume~30, pages 5360--5370, 2017.

\bibitem[Bie91]{bienstock1991}
Daniel Bienstock.
\newblock Some provably hard crossing number problems.
\newblock {\em Discrete \& Computational Geometry}, 6(3):443--459, 1991.

\bibitem[CEE{\etalchar{+}}14]{Chimani_2014_stress}
Markus Chimani, Patrick Eades, Peter Eades, Seok-Hee Hong, Weidong Huang,
  Karsten Klein, Michael Marner, Ross~T. Smith, and Bruce~H. Thomas.
\newblock People prefer less stress and fewer crossings.
\newblock In Christian Duncan and Antonios Symvonis, editors, {\em Graph
  Drawing}, volume 8871 of {\em LNCS}, pages 523--524. Springer, 2014.

\bibitem[CGJ{\etalchar{+}}14]{Chimani2014OGDF}
Markus Chimani, Carsten Gutwenger, Michael J{\"u}nger, Gunnar~W. Klau, Karsten
  Klein, and Petra Mutzel.
\newblock The open graph drawing framework ({OGDF}).
\newblock In Roberto Tamassia, editor, {\em Handbook of Graph Drawing and
  Visualization}, chapter~17, pages 543--569. CRC Press, 2014.

\bibitem[CW16]{chimani2016ilp}
Markus Chimani and Tilo Wiedera.
\newblock An {ILP}-based proof system for the crossing number problem.
\newblock In {\em 24th Annual European Symposium on Algorithms (ESA 2016)},
  volume~57 of {\em LIPIcs}, pages 29:1--29:13, 2016.

\bibitem[DADL{\etalchar{+}}19]{spx}
Sabin Devkota, Reyan Ahmed, Felice De~Luca, Katherine~E. Isaacs, and Stephen
  Kobourov.
\newblock {Stress-Plus-X} ({SPX}) graph layout.
\newblock In {\em Graph Drawing and Network Visualization: 27th International
  Symposium, GD 2019, Proceedings}, volume 11904 of {\em Lecture Notes in
  Computer Science}, pages 291--304, Cham, 2019. Springer.

\bibitem[DGL{\etalchar{+}}97]{rome_graphs}
G.~{Di Battista}, A.~Garg, G.~Liotta, R.~Tamassia, E.~Tassinari, and F.~Vargiu.
\newblock An experimental comparison of four graph drawing algorithms.
\newblock {\em Computational Geometry: Theory and Applications},
  7(5--6):303--325, 1997.

\bibitem[DH11]{Davis_Hu_ufl_2009}
Timothy~A. Davis and Yifan Hu.
\newblock The {University of Florida} sparse matrix collection.
\newblock {\em ACM Transactions on Mathematical Software}, 38(1), 2011.

\bibitem[Ead84]{Eades_1984}
P.~Eades.
\newblock A heuristic for graph drawing.
\newblock {\em Congressus Numerantium}, 42:149--160, 1984.

\bibitem[EBN{\etalchar{+}}24]{eijkelboom2024vfm}
Floor Eijkelboom, Grigory Bartosh, Christian~Andersson Naesseth, Max Welling,
  and Jan-Willem van~de Meent.
\newblock Variational flow matching for graph generation.
\newblock In {\em Advances in Neural Information Processing Systems},
  volume~37, pages 11735--11764, 2024.
\newblock arXiv:2406.04843.

\bibitem[EGK{\etalchar{+}}01]{ellson2001graphviz}
John Ellson, Emden~R. Gansner, Eleftherios Koutsofios, Stephen~C. North, and
  Gordon Woodhull.
\newblock Graphviz -- open source graph drawing tools.
\newblock In {\em Graph Drawing (GD 2001)}, volume 2265 of {\em LNCS}, pages
  483--484, 2001.

\bibitem[FR91]{Fruchterman_Reingold_1991}
T.~M.~J. Fruchterman and E.~M. Reingold.
\newblock Graph drawing by force-directed placement.
\newblock {\em Software: Practice and Experience}, 21(11):1129--1164, 1991.

\bibitem[GJ83]{garey1983crossing}
Michael~R. Garey and David~S. Johnson.
\newblock Crossing number is {NP}-complete.
\newblock {\em SIAM Journal on Algebraic and Discrete Methods}, 4(3):312--316,
  1983.

\bibitem[GKN05]{neato}
Emden~R. Gansner, Yehuda Koren, and Stephen~C. North.
\newblock Graph drawing by stress majorization.
\newblock In {\em Graph Drawing}, volume 3383 of {\em Lecture Notes in Computer
  Science}, pages 239--250, Berlin, Heidelberg, 2005. Springer.

\bibitem[GLA{\etalchar{+}}21]{dnn}
Loann Giovannangeli, Frederic Lalanne, David Auber, Romain Giot, and Romain
  Bourqui.
\newblock Deep neural network for drawing networks, ${(DNN)}^2$.
\newblock In {\em Proc. Springer International Symposium on Graph Drawing and
  Network Visualization}, volume 12868 of {\em LNCS}, pages 375--390, 2021.

\bibitem[HJA20]{Ho_2020_denoising_diffusion}
Jonathan Ho, Ajay Jain, and Pieter Abbeel.
\newblock Denoising diffusion probabilistic models.
\newblock In {\em Advances in Neural Information Processing Systems},
  volume~33, pages 6840--6851, 2020.

\bibitem[HQL{\etalchar{+}}26]{hu2026elf}
Keya Hu, Linlu Qiu, Yiyang Lu, Hanhong Zhao, Tianhong Li, Yoon Kim, Jacob
  Andreas, and Kaiming He.
\newblock {ELF}: Embedded language flows, 2026.
\newblock arXiv:2605.10938.

\bibitem[Hu05]{sfdp}
Yifan Hu.
\newblock Efficient and high quality force-directed graph drawing.
\newblock {\em Mathematica Journal}, 10(1):37--71, 2005.

\bibitem[JLB{\etalchar{+}}26]{ji2026_datafly}
Ya~Ji, Xuefeng Li, Timo Brand, Jacob Miller, Peng Zhang, Stephen Kobourov, and
  Yifan Hu.
\newblock Visualizing high-dimensional graph embeddings via informed multi-view
  projections, 2026.
\newblock arXiv:2606.31119.

\bibitem[KK89]{kamada_kawai_1989}
Tomihisa Kamada and Satoru Kawai.
\newblock An algorithm for drawing general undirected graphs.
\newblock {\em Information Processing Letters}, 31(1):7--15, 1989.

\bibitem[KM20]{kwon-ma-2020}
Oh-Hyun Kwon and Kwan-Liu Ma.
\newblock A deep generative model for graph layout.
\newblock {\em IEEE Transactions on Visualization and Computer Graphics},
  26(1):665--675, 2020.

\bibitem[Kob13]{kobourov_2013}
Stephen~G. Kobourov.
\newblock Force-directed drawing algorithms.
\newblock In Roberto Tamassia, editor, {\em Handbook of Graph Drawing and
  Visualization}, pages 383--408. CRC Press, Boca Raton, 2013.

\bibitem[LCBH{\etalchar{+}}23]{lipman2023flow}
Yaron Lipman, Ricky T.~Q. Chen, Heli Ben-Hamu, Maximilian Nickel, and Matt Le.
\newblock Flow matching for generative modeling.
\newblock In {\em International Conference on Learning Representations (ICLR)},
  2023.

\bibitem[LGL23]{liu2023rectified}
Xingchao Liu, Chengyue Gong, and Qiang Liu.
\newblock Flow straight and fast: Learning to generate and transfer data with
  rectified flow.
\newblock In {\em International Conference on Learning Representations (ICLR)},
  2023.

\bibitem[Lu26]{lu2026digfm}
Susie Lu.
\newblock Efficient graph generation via direct prediction and flow matching.
\newblock In {\em NeurIPS 2026 Workshop on Bridging Optimal Transport, Learning
  and Structured Data: Toward Geometric Distributional Learning}, 2026.
\newblock arXiv:2610.05397.

\bibitem[LZM{\etalchar{+}}26]{li2026diffgraph}
Guozheng Li, Xudong Zhou, Haotian Mi, Jinyuan Liang, Min Lu, Yi~Chen, Takayuki
  Itoh, and Chi~Harold Liu.
\newblock {DiffGraph}: Controllable graph layout generation with diffusion
  models.
\newblock In {\em IEEE Transactions on Visualization and Computer Graphics
  (Proc. IEEE VIS 2026)}, 2026.
\newblock To appear.

\bibitem[MB26]{miele2026sbd}
Andrea Miele and Ilija Bogunovic.
\newblock Spherical {Brownian} diffusion for language modeling, 2026.
\newblock arXiv:2610.35101.

\bibitem[MHS{\etalchar{+}}22]{meng2022sdedit}
Chenlin Meng, Yutong He, Yang Song, Jiaming Song, Jiajun Wu, Jun-Yan Zhu, and
  Stefano Ermon.
\newblock {SDEdit}: Guided image synthesis and editing with stochastic
  differential equations.
\newblock In {\em International Conference on Learning Representations (ICLR)},
  2022.

\bibitem[Pur97]{Purchase_1997}
H.~C. Purchase.
\newblock Which aesthetic has the greatest effect on human understanding?
\newblock In Giuseppe DiBattista, editor, {\em Proc. 5th Intl. Symp. Graph
  Drawing (GD '97)}, volume 1353 of {\em LNCS}, pages 248--261, Berlin,
  Heidelberg, 1997. Springer-Verlag.

\bibitem[RR19]{radermacher2019geometric}
Marcel Radermacher and Ignaz Rutter.
\newblock Geometric crossing-minimization -- a scalable randomized approach.
\newblock In {\em 27th Annual European Symposium on Algorithms (ESA 2019)},
  volume 144 of {\em LIPIcs}, pages 76:1--76:16, 2019.
\newblock arXiv:1907.01243.

\bibitem[RRRW19]{xing-heuristic}
Marcel Radermacher, Klara Reichard, Ignaz Rutter, and Dorothea Wagner.
\newblock Geometric heuristics for rectilinear crossing minimization.
\newblock {\em ACM J. Exp. Algorithmics}, 24(1), 2019.

\bibitem[SHW21]{satorras_21_equivariant_gcn}
V\'{\i}ctor~Garcia Satorras, Emiel Hoogeboom, and Max Welling.
\newblock E(n) equivariant graph neural networks.
\newblock In Marina Meila and Tong Zhang, editors, {\em Proceedings of the 38th
  International Conference on Machine Learning}, volume 139 of {\em Proceedings
  of Machine Learning Research}, pages 9323--9332. PMLR, 18--24 Jul 2021.

\bibitem[TCG24]{gnn-gd}
Matteo Tiezzi, Gabriele Ciravegna, and Marco Gori.
\newblock Graph neural networks for graph drawing.
\newblock {\em IEEE Transactions on Neural Networks and Learning Systems},
  35(4):4668--4681, 2024.

\bibitem[Tut63]{tutte_1963}
W.~T. Tutte.
\newblock How to draw a graph.
\newblock {\em Proceedings of the London Mathematical Society},
  s3-13(1):743--767, 1963.

\bibitem[WJW{\etalchar{+}}20]{deep-drawing}
Yong Wang, Zhihua Jin, Qianwen Wang, Weiwei Cui, Tengfei Ma, and Huamin Qu.
\newblock {DeepDrawing}: A deep learning approach to graph drawing.
\newblock {\em IEEE Transactions on Visualization and Computer Graphics},
  26(1):676--686, 2020.

\bibitem[WYHS21]{deepgd}
Xiaoqi Wang, Kevin Yen, Yifan Hu, and Han-Wei Shen.
\newblock Deep{GD}: A deep learning framework for graph drawing using {GNN}.
\newblock {\em IEEE Computer Graphics and Applications}, 41(5):32--44, 2021.

\bibitem[WYHS24]{wang_2023_smartgd}
Xiaoqi Wang, Kevin Yen, Yifan Hu, and Han-Wei Shen.
\newblock Smart{GD}: A {GAN}-based graph drawing framework for diverse
  aesthetic goals.
\newblock {\em IEEE Transactions on Visualization and Computer Graphics},
  30(8):5666--5678, 2024.

\bibitem[ZPG19]{zheng-gd2}
J.~X. Zheng, S.~Pawar, and D.~F.~M. Goodman.
\newblock Graph drawing by stochastic gradient descent.
\newblock {\em IEEE Transactions on Visualization and Computer Graphics},
  25(9):2738--2748, sep 2019.

\end{thebibliography}

\twocolumn[{\begin{center}{\Large\bfseries Supplemental Material}\end{center}\vspace{1em}}]
\setcounter{section}{0}\setcounter{table}{0}\setcounter{figure}{0}\setcounter{equation}{0}
\renewcommand{\thesection}{S\arabic{section}}
\renewcommand{\thetable}{S\arabic{table}}
\renewcommand{\thefigure}{S\arabic{figure}}
\renewcommand{\theequation}{S\arabic{equation}}
\renewcommand{\theHsection}{S\arabic{section}}
\renewcommand{\theHtable}{S\arabic{table}}
\renewcommand{\theHfigure}{S\arabic{figure}}
\renewcommand{\theHequation}{S\arabic{equation}}
This supplement gives details of the datasets (\cref{S-sec:datasets}), the model architecture (\cref{S-sec:arch}), the vertex-moving search that builds the targets (\cref{S-sec:vmove}) and the consistency experiment of \cref{sec:consistency} (\cref{S-sec:consistency}). It also reports search at matched compute (\cref{S-sec:pareto}), results on Large-Val (\cref{S-sec:largeval}), per-graph results on the named RR graphs (\cref{S-sec:rrnamed}), the effect of the sampling seed and the number of sampling steps (\cref{S-sec:seeds}), and the stress polish (\cref{S-sec:polish}). Section, table, figure and equation numbers with an ``S'' refer to this supplement.

\section{Datasets}
\label{S-sec:datasets}

\subsection{Rome}
\label{S-sec:rome}
The Rome graphs~\cite{rome_graphs} are 11,534 graphs with 10--110 nodes, generated from 112 seed graphs; 10,756 of
them have a known exact crossing number~\cite{chimani2016ilp}. We use two splits.

\emph{SmartGD split, Rome${}_S$.} XFlow-cross and all results in Section~\ref{sec:rq3} use the split of
SmartGD~\cite{wang_2023_smartgd}: 10,000 training, 531 validation and 1,000 test graphs. The test graphs have 30--49
or 90--109 nodes; 783 of them have a known exact crossing number, and the share of drawings at the optimum is taken
over these.

\emph{Earlier split, Rome${}_R$.} The analyses in Sections~\ref{sec:rq1} and~\ref{sec:rq2} use an earlier split of the
10,756 graphs with a known exact crossing number into 9,656 training, 100 validation and 1,000 test graphs; Targets A
and B and the models of these sections are built on its training graphs.
Two exceptions: the stress models of Section~\ref{sec:rq1} (Tables~\ref{tab:stress} and~\ref{tab:ood}) are trained on 5,000 and tested on 1,000 graphs of a third random split, whose training graphs include none of the Rome${}_R$ test graphs; and the consistency experiment (\cref{S-sec:consistency}) uses Rome${}_S$ test graphs.

The two test sets differ in size: neato draws the SmartGD test graphs with 57.22 crossings on average (exact crossing
number 7.84) and the earlier split's test graphs with 25.23 (exact crossing number 5.29), so crossing counts from
Sections~\ref{sec:rq1} and~\ref{sec:rq2} are not comparable with those of Section~\ref{sec:rq3}.

All Rome graphs derive from the same 112 seed graphs, so any Rome test split measures in-distribution performance
only.

\subsection{Diverse, Diverse-Aug, Large-Val, Large-Test and Mixed Datasets}
\label{S-sec:diverse}
Rome graphs are small (at most 110 nodes), have small diameter (at most 29) and are sparse (about 1.3 edges per
node). The out-of-domain benchmark graphs (\cref{S-sec:rr}) have up to 7,920 nodes, diameters up to 156 and up to 14
edges per node. Diverse widens the training distribution on all three axes: it has 20--1,000 nodes
per graph, diameters up to about 85 and 1--15 edges per node. It has three sources; every graph is first generated
and laid out with \texttt{neato} (default settings), then the pool is split, and only then is any model trained.

\emph{Synthetic graphs.} 3,250 graphs from the ten families of \cref{S-tab:families}, 325 per family, generated with
NetworkX (and SciPy for the Delaunay meshes). The target size $n$ is drawn log-uniformly from 20--1,000, so every
octave of sizes is equally represented. We keep the largest connected component and redraw the graph if it has
fewer than 20 nodes, and shuffle the node labels. Random regular graphs never have exactly 1,000 nodes, so none
coincides in size with the regular graphs of the benchmark.

\begin{table}[t]
\centering\small
\caption{The ten synthetic families of Diverse (325 graphs each). $U(a,b)$: uniform on $[a,b]$.}
\label{S-tab:families}
\resizebox{\columnwidth}{!}{%
\begin{tabular}{lll}
\toprule
Family & Generator & Random parameters \\
\midrule
geometric & random geometric graph & mean degree $U(4,10)$ \\
Delaunay & Delaunay triangulation of random points & aspect $U(1,6)$; delete $U(0,30)\%$ of edges \\
lattice & square, triangular or hexagonal grid & aspect $U(1,5)$; delete $U(0,15)\%$ of edges \\
tree & random tree (Pr\"ufer sequence) & extra random edges $U(0,10)\%$ of $n$ \\
Barab\'asi--Albert & preferential attachment & $m \in \{1,\dots,8\}$ \\
Watts--Strogatz & small world & $k \in \{4,6,8\}$; rewiring $U(0.01,0.3)$ \\
Erd\H{o}s--R\'enyi & $G(n,m)$ & mean degree log-uniform on $[2.5,30]$ \\
block model & stochastic block model & 2--10 blocks; degree $U(4,24)$; in/out ratio $U(10,50)$ \\
regular & random regular graph & $k \in \{3,\dots,26\}$ \\
power-law cluster & Holme--Kim & $m \in \{1,2,3\}$; triangle prob.\ $U(0.1,0.8)$ \\
\bottomrule
\end{tabular}%
}
\end{table}

\emph{SuiteSparse matrices.} From the SuiteSparse Matrix Collection~\cite{Davis_Hu_ufl_2009} we take the square
matrices with 20--1,200 rows (470), and exclude every group or matrix family that contains a benchmark graph
(Newman, Arenas, JGD\_Homology, Pothen, and the HB families \texttt{bcss*}, \texttt{*\_bus}, \texttt{mahindas},
\texttt{eris*}) as well as series of nearly identical matrices (\emph{Subsequent}, \emph{Duplicate} or
\emph{Sequence} kinds), which leaves 361. Each matrix is read as an undirected adjacency pattern without self-loops;
we keep the largest connected component if it has 20--1,000 nodes. This gives 282 graphs (339 matrices could be
downloaded), with median 399 nodes, diameter up to 500 and up to 53.5 edges per node.

\emph{Rome.} 2,000 Rome training graphs and 50 Rome validation graphs, drawn at random from the Rome graphs outside
the earlier split's validation and test graphs (1,901 of them are in its training set; the other 149 have no known
crossing number).

\emph{Split.} Synthetic graphs are split within each family: 300 training, 5 validation and 20 test graphs.
SuiteSparse graphs are split by matrix family (group plus name up to the first digit, e.g.\ all \texttt{HB/can\_*}),
so that near-identical matrices never straddle training and test: 259 training, 5 validation and 18 test graphs. A
family goes to validation or test only if it fits in the remaining quota, so that no single large family fills a
split. In total there are 5,259 training graphs (2,000 Rome, 3,000 synthetic, 259 SuiteSparse), 105 validation graphs
(50, 50, 5) and 218 test graphs (200 synthetic and 18 SuiteSparse; the Rome test graphs are kept separate). The
synthetic test graphs come from families seen in training, so they measure coverage of the training distribution;
the SuiteSparse test graphs come from unseen matrix families.

\emph{Diverse-Aug.} Following DiffGraph, we enlarge the training split (never validation or test) to about 58,000
graphs. Each training graph gets eight variants with a random 5--40\% of its nodes removed (largest connected
component, retried until it keeps at least half of the remaining nodes and at least 10 nodes) and two variants grown
by node duplication, as in the duplication--divergence model: 20--100\% more nodes (at most 2,000), each new node
copying the edges of a random existing node, each edge kept with a per-graph probability $U(0.3,0.8)$ and at least
one kept. Every variant gets its own \texttt{neato} layout.

\emph{Large-Val.} The 100 Diverse graphs with more than 300 nodes that we hold out for choosing
sampler settings on large graphs: the 18 such graphs of the validation split and 82 held-out graphs of the training
split. Their mean \texttt{neato} crossing count is 605K.

\emph{Large-Test.} The 78 graphs of Diverse's test split with more than 300 nodes (309--999 nodes; 63 synthetic graphs and 15 SuiteSparse matrices), used only for testing. Their mean \texttt{neato} crossing count is 242K.

\emph{Mixed.} The training set of XFlow-cross: the 10,000 Rome${}_S$ training graphs, the 3,177 Diverse
training graphs that are neither Rome graphs nor in Large-Val, and 4,000 Diverse-Aug variants with
300--1,000 nodes of these graphs. Graphs with more than 10,000 edges (77) are left out, which leaves 17,100 graphs.
No validation or test graph of any collection, nor any variant of one, is used for training.

\subsection{RR benchmark}
\label{S-sec:rr}
The benchmark of Radermacher and Rutter~\cite{radermacher2019geometric} has 16 named graphs, 9 from the 10th DIMACS
Implementation Challenge and 7 from the SuiteSparse Matrix Collection, with 102--7,920 nodes. As in their paper, we
keep the largest connected component and then repeatedly delete degree-1 nodes; with this, the node and edge counts
of all 16 graphs match their Table 1 exactly. Their random regular graphs are not published, so we generated new
instances of the same kind: 25 random $k$-regular graphs with 1,000 nodes for each of $k = 3, 6, 9$ (NetworkX,
seeds 0--24). No benchmark graph or variant of one is used for training or for choosing settings.

\section{Architecture details}
\label{S-sec:arch}

XFlow's denoiser maps a noisy layout $x_t \in \mathbb{R}^{n\times 2}$ and noise level $t$ to a predicted layout
$\hat x$. It keeps two quantities per node: a hidden state $h_i \in \mathbb{R}^{240}$, which never sees absolute
positions, and a position $x_i \in \mathbb{R}^2$. Both are updated in each of the 8 layers.

\emph{Input.} Each node has 9 structural features: its random-walk return probabilities $(P^k)_{ii}$, $k=1,\dots,8$,
with $P = D^{-1}A$, and $\log(1+\deg_i)$; XFlow-Pivot adds a tenth that flags pivots. An MLP maps them to $h^{(0)}_i \in \mathbb{R}^{240}$. The positions start at the noisy layout, $x^{(0)} = x_t$. 

\emph{Layer $\ell = 1,\dots,8$.} Let $h = h^{(\ell-1)}$ and $x = x^{(\ell-1)}$. With $H = 8$ heads of size $d_h = 30$
and layer norms $\mathrm{LN}$ (each with its own parameters), a layer computes:
\begin{align}
  h_i &\leftarrow h_i + W_t\,\tau(t), \\
  (q^h_i, k^h_i, v^h_i) &= W_{qkv}\,\mathrm{LN}(h_i), \\
  a^h_{ij} &= \operatorname{softmax}_{j}\Bigl( \frac{(q^h_i)^{\top} k^h_j}{\sqrt{d_h}}
      + b^h_{\mathrm{spd}}(d_{ij}) + b^h_{\mathrm{dist}}\bigl(\|x_i - x_j\|\bigr) \Bigr), \\
  h_i &\leftarrow h_i + W_o\,\bigl[\textstyle\sum_j a^h_{ij}\, v^h_j\bigr]_{h=1}^{H}, \\
  (u^h_i, w^h_i) &= W_{uw}\,\mathrm{LN}(h_i), \qquad
  c^h_{ij} = (u^h_i)^{\top} w^h_j / \sqrt{d_h}, \\
  x^{(\ell)}_i &= x_i + \sum_j \sum_{h=1}^{H} a^h_{ij}\, c^h_{ij}\, \frac{x_i - x_j}{\|x_i - x_j\| + 1}, \\
  h^{(\ell)}_i &= h_i + \mathrm{FFN}\bigl(\mathrm{LN}(h_i)\bigr).
\end{align}
The sums over $j$ run over all nodes of the graph (for XFlow-Pivot, over the graph's neighbors of $i$ and the
pivots). $[\cdot]_{h=1}^{H}$ concatenates the heads, and $\mathrm{FFN}$ is Linear($240\to960$), SiLU,
Linear($960\to240$). $W_t$, $W_{qkv}$, $W_o$, $W_{uw}$ are learned linear maps, separate in each layer.

\emph{Attention biases.} $d_{ij}$ is the shortest-path distance between $i$ and $j$. $b^h_{\mathrm{spd}}$ is a learned
table with one entry per head and distance: exact distances up to 16, then 8 buckets per doubling of the distance up to
1,024. It is initialized to $-\log d_{ij}$, so that attention starts proportional to $1/d_{ij}$. $b^h_{\mathrm{dist}}$
is a small MLP (Linear $1\to16$, SiLU, Linear $16\to H$) on the distance between the current positions. Its last layer
is initialized to zero, so it has no effect at the start of training.

\emph{Roles of $a$ and $c$.} The attention weight $a^h_{ij} \ge 0$ selects the nodes that
$i$ responds to. The step $c^h_{ij}$ is signed and unnormalized: $c > 0$ moves $i$ away from $j$, $c < 0$ toward it.
It is computed from the hidden states after this layer's attention update. 
$c$ depends on positions only
through $\|x_i - x_j\|$, and positions move only along $x_i - x_j$. Rotating or reflecting $x_t$ therefore rotates or
reflects every $x^{(\ell)}$ in the same way, so the model is E(2)-equivariant.

\emph{Output.} The prediction is the last layer's positions, centered
$\hat x_i = x^{(8)}_i - \frac{1}{n}\sum_k x^{(8)}_k$. The hidden states are not projected to coordinates.
In total the model has 7.1M parameters.

\section{Vertex-moving search}
\label{S-sec:vmove}

Targets A and B (Section~\ref{sec:learnx} of the paper) are built with a straight-line crossing-minimization search
in the style of Radermacher and Rutter~\cite{radermacher2019geometric}. It repeatedly moves one vertex at a time to a
position that lowers the crossings of its incident edges.

\emph{One pass.} Let $\ell$ be the median edge length of the current drawing. A pass visits the vertices in random
order. For each vertex $v$ it samples 640 candidate positions, plus $v$'s current position:
\begin{itemize}
\item 256 uniformly in the drawing's bounding box, enlarged by $\ell$ on each side (long moves);
\item 256 near the neighbours of $v$: a random neighbour's position plus Gaussian noise of standard deviation $\ell$;
\item 128 near the current position, plus Gaussian noise of standard deviation $0.1\ell$, $0.3\ell$, $\ell$ and $3\ell$
  (32 each).
\end{itemize}
Candidates closer than $0.3\ell$ to another vertex are discarded, thereby keeping vertices apart. Each remaining candidate $p$
is scored by the number of proper crossings between $v$'s edges, with $v$ placed at $p$, and the edges not incident to
$v$ (edges that share an endpoint are not counted). Ties are broken by $v$'s stress terms, $\sum_{u\ne v}
\bigl((\alpha\|p - x_u\| - d_{uv})/d_{uv}\bigr)^2$, where $d$ is the shortest-path distance and $\alpha$ the optimal
scale of the current drawing. The vertex moves to the best candidate if this lowers its crossings, or keeps them
equal and lowers its stress; otherwise it stays.

\emph{Stopping.} Passes repeat until the drawing reaches the graph's exact crossing number (and is therefore
optimal), until 3 passes in a row fail to lower the total number of crossings, or until a maximum number of passes.

\emph{Starts and polish.} The search runs from several starting drawings and keeps the one with the fewest crossings.
It then runs a fixed number of further passes on that drawing (polish passes), without the early stop. These rarely
remove crossings and mostly move vertices to positions with the same crossings and lower stress.

\begin{itemize}
\item \emph{Target A:} starts from OGDF's planarization drawing with the missing edges added back, from 3 straight-line
  drawings of random maximal planar subgraphs, and from 2 copies of the planarization drawing with every vertex
  jittered by Gaussian noise of 0.15 times the drawing's spread; at most 40 passes per start; 5 polish passes.
\item \emph{Target B:} starts from the neato layout and from 2 copies with every vertex jittered by 0.05 times the
  layout's spread; at most 60 passes per start; 10 polish passes.
\end{itemize}
All starts are skipped once one of them reaches the exact crossing number.

\emph{Cost.} Scoring a candidate takes $O(\deg(v)\, m)$ segment tests. A pass therefore costs about $640 \cdot 2m \cdot m$ tests,
serially on one CPU core. 

\section{The consistency experiment}
\label{S-sec:consistency}

\emph{Graphs.} We draw 200 graphs at random from the 1,000 Rome${}_S$ test graphs. For each, we make
five variants by adding one edge between a node pair chosen uniformly among the pairs at graph distance exactly 2
(three variants) or 3 (two variants). An edge at distance 2 closes a triangle, the smallest possible change to the
graph metric; an edge at distance 3 is a slightly larger change. Adding an edge keeps the node set and the
connectivity, so drawings of a graph and its variant can be compared node by node.

\emph{Methods.} Each original graph and each variant is drawn with every method of the paper:
\texttt{neato} and \texttt{sfdp} (Graphviz, \texttt{sfdp} with \texttt{-Goverlap=true}, as for the training targets)
and Targets A, B and C, built exactly as the training targets of Section~\ref{sec:learnx}. Each original graph is
also drawn with two more random seeds: for \texttt{neato} and \texttt{sfdp}, the seed sets Graphviz's random start;
for Targets A and B, it sets the randomness of the vertex-moving search, while the layouts they start from
(OGDF's planarization and \texttt{neato} with its default start) stay the same. Target C has no randomness given the graph: it
starts from \texttt{neato}'s default layout and SigmoidX gradient descent is deterministic.

\emph{Distance.} Let $X, Y \in \mathbb{R}^{n\times 2}$ be two drawings of the same nodes, and $\bar X$, $\bar Y$ the
drawings translated so that their centroid is at the origin. The Procrustes distance is the residual of the best
similarity transformation of $X$ onto $Y$, relative to the size of $Y$:
\begin{equation}
d(X, Y) = \min_{s \ge 0,\; R \in O(2),\; t \in \mathbb{R}^2}
\frac{\| s X R + \mathbf{1} t^\top - Y \|_F}{\| \bar Y \|_F}
= \sqrt{1 - \frac{(\sigma_1 + \sigma_2)^2}{\|\bar X\|_F^2 \, \|\bar Y\|_F^2}},
\label{S-eq:procrustes}
\end{equation}
where $O(2)$ contains the rotations and reflections, and $\sigma_1, \sigma_2$ are the singular values of
$\bar X^\top \bar Y$; the optimum is $R = UV^\top$ for the SVD $\bar X^\top \bar Y = U \Sigma V^\top$, and
$s = (\sigma_1 + \sigma_2)/\|\bar X\|_F^2$. The closed form shows that $d$ is symmetric in $X$ and $Y$, is 0 exactly
when one drawing is a similarity transform of the other, and is at most 1 (the value for $s = 0$), which it reaches
when the two drawings are uncorrelated ($\bar X^\top \bar Y = 0$). The quantity under the square root is the
\emph{disparity} returned by SciPy's \texttt{scipy.spatial.procrustes}, which our code calls; we report its square
root because it is a distance, proportional to how far the nodes move rather than to its square.

\emph{Results.} \Cref{S-tab:consistency} gives the quartiles over all pairs. Two observations complement the paper.
First, for \texttt{neato} most reseeds give the same drawing (median 0.004), but about a quarter of the graphs have a
second stress minimum that another random start finds (upper quartile 0.49). Second, for \texttt{sfdp} and Target B the
change after an added edge is about as large as the change after a reseed: much of their sensitivity is a choice
between equally good drawings, not a response to the edit. Target A is the only method whose drawing changes far
more after an added edge (median 0.77) than after a reseed (0.30), and it is equally sensitive to edges at distance 2
and 3.

\begin{table}[t]
\centering\small
\caption{Procrustes distance (\cref{S-eq:procrustes}) between the drawing of a Rome graph and that of a variant with one
added edge between nodes at graph distance 2 (600 pairs) or 3 (400 pairs), or of the same graph drawn with another
seed (400 pairs); 25th percentile, median and 75th percentile over 200 graphs.}
\label{S-tab:consistency}
\resizebox{\columnwidth}{!}{%
\begin{tabular}{lccc}
\toprule
Method & Added edge, distance 2 & Added edge, distance 3 & Other seed \\
\midrule
\texttt{neato} & 0.036 / 0.073 / 0.132 & 0.072 / 0.126 / 0.278 & 0.000 / 0.004 / 0.488 \\
\texttt{sfdp} & 0.062 / 0.142 / 0.422 & 0.103 / 0.217 / 0.446 & 0.089 / 0.180 / 0.410 \\
Target A & 0.433 / 0.768 / 0.874 & 0.579 / 0.770 / 0.881 & 0.209 / 0.299 / 0.428 \\
Target B & 0.148 / 0.252 / 0.352 & 0.200 / 0.283 / 0.397 & 0.120 / 0.217 / 0.288 \\
Target C & 0.205 / 0.309 / 0.389 & 0.252 / 0.353 / 0.434 & 0.000 / 0.000 / 0.000 \\
\bottomrule
\end{tabular}%
}
\end{table}

\section{Search at matched compute}
\label{S-sec:pareto}

\begin{figure*}[t]
\centering
\includegraphics[width=\linewidth]{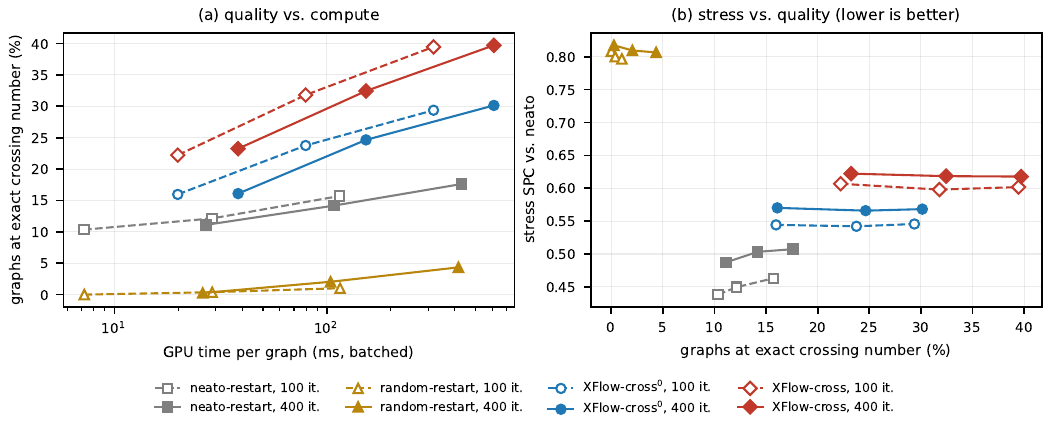}
\caption{Rome${}_S$ test graphs (1,000): best of $k \in \{1, 4, 16\}$ starts refined by 100 (open markers) or 400 (filled)
SigmoidX iterations. (a) Share of graphs whose drawing reaches the exact crossing number, against batched GPU time
per graph; restart rows also need a neato run. (b) Stress SPC against neato for the same runs. The model dominates at
every budget; it pays for its fewer crossings with somewhat higher stress than perturbed restarts.}
\label{S-fig:pareto}
\end{figure*}

\Cref{S-fig:pareto} plots the runs of \cref{tab:rome} against their batched GPU time per graph. In (a), XFlow-cross
reaches the exact crossing number more often than every classical start at every budget: with one sample and no
refinement (12.8\,ms) it already matches the best of 16 neato restarts with 400 SigmoidX iterations each (20\% vs.\
18\%, at 432\,ms), and with 16 samples and 100 iterations (318\,ms) it reaches 39\%. Neato restarts gain little from
more starts (11\%, 14\% and 18\% for $k = 1$, 4 and 16), and random starts stay below 5\% even with 16 starts.
Raising SigmoidX from 100 to 400 iterations roughly doubles the time of the learned starts but adds only one point
(39\% to 40\%), so the starts, not the refinement budget, decide the result. Expert iteration shifts the whole curve
up: XFlow-cross$^0$ reaches 29\% at the same 318\,ms. In (b), the learned starts pay for their fewer crossings with
somewhat higher stress than neato restarts (\sspc{} about +0.60 vs.\ +0.51 against neato), but far lower stress than
random starts (+0.81).

\section{Results on Large-Val}
\label{S-sec:largeval}

Large-Val is the validation set on which we chose the sampler settings for large graphs (re-noising time $t_0$, chains and rounds, and $k = 16$; \cref{sec:renoise}), so the main paper reports Large-Test instead. \Cref{S-tab:largeval} gives the Large-Val results with the protocol of \cref{tab:large}. The margins over neato-restart are slightly larger than on Large-Test ($-$0.230 vs.\ $-$0.203 for best of 16, $-$0.245 vs.\ $-$0.221 for re-noising), as expected for the set used to choose $t_0$. Re-noising with $t_0 = 0.5$ beats $t_0 = 0.3$ (54 wins, 34 losses, $p = 0.04$) and $t_0 = 0.8$ (52/35, $p = 0.09$), although all three beat neato-restart by a similar margin. Across the expert-iteration rounds of \cref{tab:iter}, the re-noising chains reach \xspc{} against neato of $-$0.511, $-$0.507, $-$0.526, $-$0.528 and $-$0.519 (rounds 0--4).

\begin{table}[htb]
\centering\small
\caption{Large-Val (100 graphs): 16 SigmoidX runs of 100 iterations per graph (random-restart: 4). \xspc{} and \sspc{} against plain neato; \xspc{} and wins/losses (W/L) against neato-restart (noise 0.1).}
\label{S-tab:largeval}
\setlength{\tabcolsep}{3.5pt}
\begin{tabular}{lcccc}
\toprule
 & \multicolumn{2}{c}{vs.\ neato} & \multicolumn{2}{c}{vs.\ neato-restart} \\
\cmidrule(lr){2-3}\cmidrule(lr){4-5}
Starts & \xspc\,$\downarrow$ & \sspc\,$\downarrow$ & \xspc\,$\downarrow$ & W/L \\
\midrule
neato-restart & $-$0.443 & +0.36 & 0.000 & -- \\
random-restart ($k$=4) & +0.323 & +0.75 & +0.618 & 1/99 \\
XFlow-cross, best of 16 & $-$0.518 & +0.47 & $-$0.230 & 83/10 \\
XFlow-cross, re-noise & $-$0.519 & +0.50 & $-$0.245 & 89/7 \\
\bottomrule
\end{tabular}
\end{table}

\section{Per-graph results on the named RR graphs}
\label{S-sec:rrnamed}

\Cref{S-tab:rr} gives the crossings of every method on the 16 named graphs of Radermacher and
Rutter's benchmark~\cite{radermacher2019geometric}, together with their published results. Best of 16 XFlow-cross
samples refined by SigmoidX beats their best heuristic, S512, on 15 graphs, and re-noising chains beat it on all 16
(geometric mean 0.697 relative to S512, i.e., 30\% fewer crossings).
\begin{table*}[htb]
\centering\footnotesize
\caption{Crossings on the 16 named RR graphs. SigmoidX from: best of $k$ runs of 100 iterations.
Published~\cite{radermacher2019geometric}: their stress layout and their heuristics S512 and S0;
SGD${}^2$~\cite{ahmed2022multicriteria}: our run, not refined. Bold: fewest crossings in the row.}
\label{S-tab:rr}
\setlength{\tabcolsep}{3pt}
\begin{tabular}{lrrrrrrrrrrrr}
\toprule
 & & & & \multicolumn{5}{c}{SigmoidX from} & \multicolumn{4}{c}{published} \\
\cmidrule(lr){5-9}\cmidrule(lr){10-13}
Graph & $n$ & $m$ & neato &  \shortstack{neato\\$k$=1} & \shortstack{neato-\\$k$=16} & \shortstack{random-\\$k$=16} & \shortstack{XFlow\\$k$=16} & \shortstack{XFlow\\$k$=16 re-noise} & Stress & S512 & S0 & {\color{black}SGD${}^2$} \\
\midrule
adjnoun & 102 & 415 & 6,883 & 4,266 & 4,006 & 4,513 & 3,855 & \textbf{3,750} & 6,576 & 3,775 & 4,468 & {\color{black}5,112} \\
football & 115 & 613 & 6,765 & 3,476 & 3,274 & 4,081 & 3,204 & \textbf{3,159} & 6,865 & 3,568 & 4,030 & {\color{black}4,376} \\
netscience & 352 & 887 & 1,879 & 507 & 492 & 2,285 & 479 & \textbf{375} & 1,724 & 583 & 814 & {\color{black}1,519} \\
celegans\_metabolic & 445 & 2,017 & 119,664 & 56,518 & 54,958 & 69,318 & 52,100 & \textbf{49,725} & 113,117 & 55,714 & 63,028 & {\color{black}91,383} \\
celegansneural & 282 & 2,133 & 132,268 & 79,945 & 78,845 & 91,119 & 75,468 & \textbf{74,553} & 128,068 & 86,641 & 90,920 & {\color{black}112,187} \\
jazz & 193 & 2,737 & 221,626 & 145,791 & 143,939 & 153,202 & \textbf{133,510} & 134,949 & 223,990 & 143,647 & 153,040 & {\color{black}191,947} \\
power & 3,353 & 5,006 & 7,696 & 2,478 & 2,458 & 137,542 & \textbf{1,820} & 1,880 & 7,622 & 6,854 & 6,293 & {\color{black}332,909} \\
email & 978 & 5,296 & 535,362 & 280,960 & 279,005 & 404,326 & 269,400 & \textbf{265,617} & 504,144 & 342,020 & 357,272 & {\color{black}533,799} \\
hep-th & 4,786 & 12,766 & 926,812 & 286,181 & 284,261 & 941,780 & 279,855 & \textbf{256,091} & 836,809 & 546,780 & 638,069 & {\color{black}3,387,303} \\
1138\_bus & 671 & 991 & 818 & 357 & 293 & 4,191 & 169 & \textbf{162} & 657 & 402 & 467 & {\color{black}1,710} \\
ch7-6-b1 & 630 & 1,243 & 64,549 & 25,297 & 24,375 & 26,374 & 23,672 & \textbf{22,873} & 64,055 & 24,928 & 26,055 & {\color{black}33,597} \\
mk9-b2 & 1,260 & 3,774 & 422,061 & 202,221 & 199,777 & 277,285 & 183,154 & \textbf{175,696} & 412,397 & 248,884 & 252,198 & {\color{black}317,566} \\
bcsstk08 & 1,055 & 5,927 & 509,479 & 251,507 & 244,502 & 367,559 & 233,402 & \textbf{231,765} & 455,069 & 342,996 & 344,644 & {\color{black}597,222} \\
mahindas & 1,258 & 7,513 & 1,519,718 & 816,347 & 809,273 & 1,008,541 & 804,491 & \textbf{780,881} & 1,463,437 & 933,247 & 1,042,787 & {\color{black}1,374,821} \\
eris1176 & 892 & 8,405 & 1,697,561 & 949,097 & 947,035 & 1,009,995 & \textbf{942,456} & 948,027 & 1,682,458 & 1,030,881 & 1,087,605 & {\color{black}1,334,346} \\
commanche\_dual & 7,920 & 11,880 & 6,645 & 4,397 & 4,397 & 1,051,012 & 3,971 & \textbf{3,541} & 6,332 & 6,239 & 6,146 & {\color{black}3,656,115} \\
\midrule
\multicolumn{3}{l}{\xspc{} relative to neato} & 0 & $-$0.509 & $-$0.524 & $-$0.065 & $-$0.557 & \textbf{$-$0.572} & $-$0.049 & $-$0.402 & $-$0.359 & {\color{black}+0.063} \\
\multicolumn{3}{l}{\xspc{} relative to S512} & +0.402 & $-$0.152 & $-$0.180 & +0.315 & $-$0.235 & \textbf{$-$0.265} & +0.371 & 0.000 & +0.077 & {\color{black}+0.458} \\
\multicolumn{3}{l}{Graphs with fewer crossings than S512} & 0 & 12 & 14 & 1 & 15 & \textbf{16} & 0 & -- & 2 & {\color{black}0} \\
\multicolumn{3}{l}{Total time, 16 graphs (s)} & 360$^\ast$ & 155 & 2,572 & 2,586 & 3,217 & 3,212 & & 23,005$^\ast$ & 14,690$^\ast$ & {\color{black}2,814$^\ast$} \\
\bottomrule
\end{tabular}

\smallskip\raggedright $^\ast$CPU: neato on Xeon E5-2680 v4; S512 and S0~\cite{radermacher2019geometric} on Xeon E5-1630
v3;%
SGD${}^2$ on AMD EPYC 9555 Turin CPU. Our times are on one V100 GPU. %
\end{table*}

\section{Impact of sampling seed and number of steps on XFlow-cross}
\label{S-sec:seeds}
Results vary little with the sampling seed: over three seeds, XFlow-cross (best of 16, 100 SigmoidX iterations)
reaches the exact crossing number on 38.7--39.5\% of the Rome${}_S$ test graphs and beats the 16 neato restarts with 400
iterations on 829--834 graphs, losing at most 3. On RR, \xspc{} against restarts is $-0.059$ to $-0.060$ for best of
16 and $-0.103$ to $-0.104$ for re-noising. Sampling with 32 instead of 8 steps lowers mean crossings only slightly
(28.28 to 27.66 for one sample, 19.69 to 19.34 for best of 16) at four times the sampling cost, so we use 8.

\section{Stress polish}
\label{S-sec:polish}
The stress polish lowers the stress of a refined layout without adding crossings. It runs 300 Adam optimization
steps (learning rate 0.05) on stress plus the SigmoidX barrier function ($T = 60$). The SigmoidX loss is normalized by the crossing count
of the starting layout so that the two terms stay comparable. Of the iterates checked every 10 steps, it keeps the one
with the fewest crossings, then the lowest stress, so it never adds crossings. In the fourth round of expert
iteration, larger graphs get 100 Adam steps.

\end{document}